\documentclass[nonacm, authorversion,preprint, sigconf]{acmart}

\usepackage{listings}

\AtBeginDocument{%
  \providecommand\BibTeX{{%
    \normalfont B\kern-0.5em{\scshape i\kern-0.25em b}\kern-0.8em\TeX}}}

\copyrightyear{2023}
\acmYear{2023}
\setcopyright{acmlicensed}\acmConference[CHI '23]{Proceedings of the 2023 CHI Conference on Human Factors in Computing Systems}{April 23--28, 2023}{Hamburg, Germany}
\acmBooktitle{Proceedings of the 2023 CHI Conference on Human Factors in Computing Systems (CHI '23), April 23--28, 2023, Hamburg, Germany}
\acmPrice{15.00}
\acmDOI{10.1145/3544548.3580754}
\acmISBN{978-1-4503-9421-5/23/04}

\begin{document}

\title[Deimos: A Grammar of Dynamic Embodied Immersive Morphs]{Deimos: A Grammar of Dynamic Embodied Immersive Visualisation Morphs and Transitions}


\author{Benjamin Lee}
\email{Benjamin.Lee@visus.uni-stuttgart.de}
\orcid{0000-0002-1171-4741}
\affiliation{%
  \institution{University of Stuttgart}
  \city{Stuttgart}
  \country{Germany}
}
\affiliation{%
  \institution{Monash University}
  \city{Melbourne}
  \country{Australia}
}

\author{Arvind Satyanarayan}
\email{arvindsatya@mit.edu}
\orcid{0000-0001-5564-635X}
\affiliation{%
  \institution{MIT CSAIL}
  \city{Cambridge}
  \country{United States}
}

\author{Maxime Cordeil}
\email{m.cordeil@uq.edu.au}
\orcid{0000-0002-9732-4874}
\affiliation{%
  \institution{University of Queensland}
  \city{Brisbane}
  \country{Australia}
}

\author{Arnaud Prouzeau}
\email{arnaud.prouzeau@inria.fr}
\orcid{0000-0003-3800-5870}
\affiliation{%
  \institution{Inria \& LaBRI (University of Bordeaux, CNRS, Bordeaux-INP)}
  \state{Bordeaux}
  \country{France}
}

\author{Bernhard Jenny}
\email{bernie.jenny@monash.edu}
\orcid{0000-0001-6101-6100}
\affiliation{%
  \institution{Monash University}
  \city{Melbourne}
  \country{Australia}
}

\author{Tim Dwyer}
\email{tim.dwyer@monash.edu}
\orcid{0000-0002-9076-9571}
\affiliation{%
  \institution{Monash University}
  \city{Melbourne}
  \country{Australia}
}

\renewcommand{\shortauthors}{Lee et al.}

\begin{abstract}
We present Deimos, a grammar for specifying dynamic embodied immersive visualisation morphs and transitions. A morph is a collection of animated transitions that are dynamically applied to immersive visualisations at runtime and is conceptually modelled as a state machine. It is comprised of state, transition, and signal specifications.
States in a morph are used to generate animation keyframes, with transitions connecting two states together. A transition is controlled by signals, which are composable data streams that can be used to enable embodied interaction techniques. Morphs allow immersive representations of data to transform and change shape through user interaction, facilitating the embodied cognition process.
We demonstrate the expressivity of Deimos in an example gallery and evaluate its usability in an expert user study of six immersive analytics researchers. Participants found the grammar to be powerful and expressive, and showed interest in drawing upon Deimos’ concepts and ideas in their own research.
\end{abstract}

\ccsdesc[500]{Human-centered computing~Visualization theory, concepts and paradigms}
\ccsdesc[500]{Human-centered computing~Visualization toolkits}
\ccsdesc[500]{Human-centered computing~Mixed / augmented reality}
\keywords{Immersive Analytics, data visualisation, animated transitions, embodied interaction, user study, grammar}

\begin{teaserfigure}
  \centering
  \includegraphics[width=0.9\textwidth]{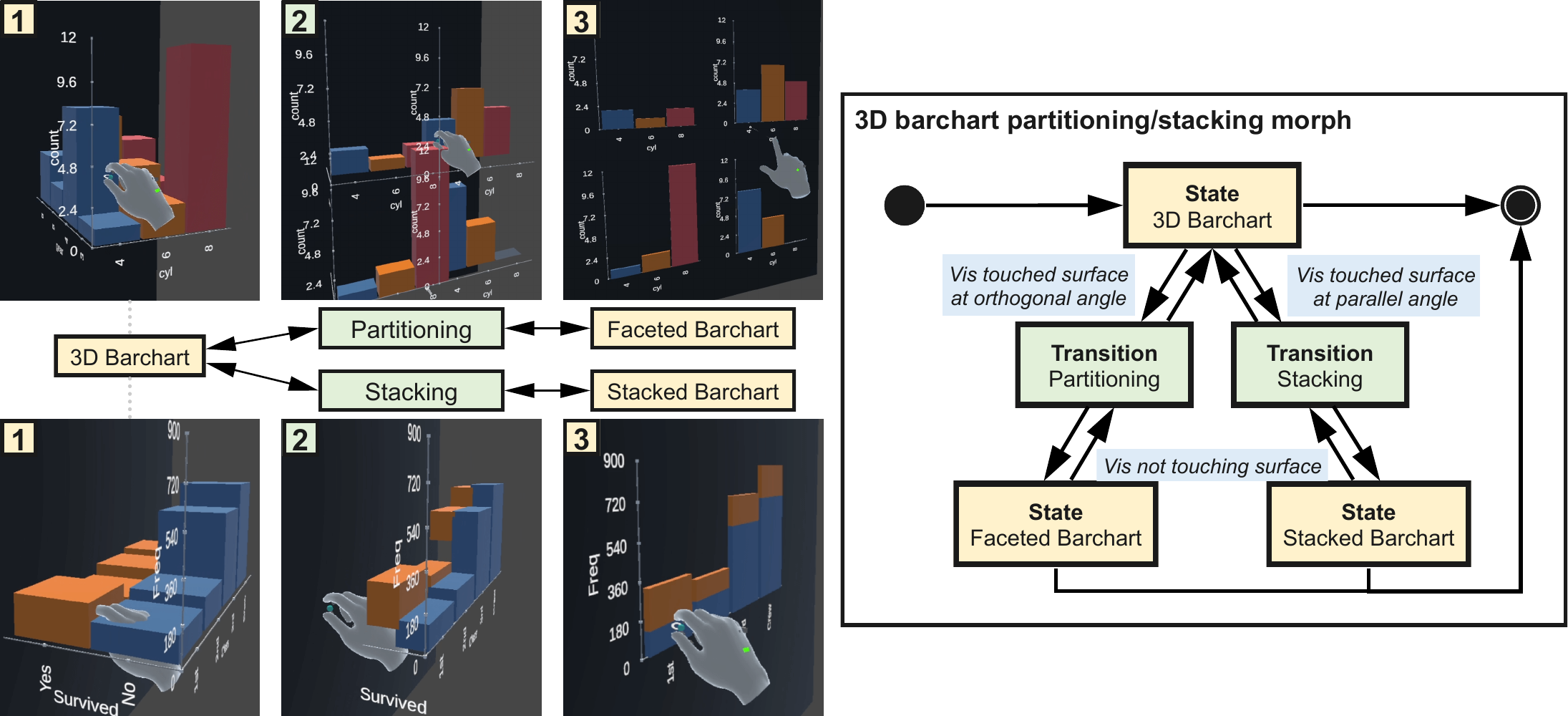}
  \caption{A morph that transforms any 3D barchart into either a faceted barchart or a stacked barchart when it touches a surface in the immersive environment, depending on its angle of intersection. Left: Still images of the animated transition. Images labelled ``1'' and ``3'' correspond to \textit{states} (keyframes) in the morph, images labelled ``2'' correspond to the \textit{transitions} in the morph. Right: State machine of the morph that corresponds with the still images via colour-coding (yellow for states, green for transitions). Blue represents \textit{signals} used to control the behaviour of transitions in the overall morph.}
  \Description{A two part image showing an example of a branching morph. The left part shows two sets of still images. The first set shows a 3D barchart being pushed into a virtual wall, it partitioning into four small multiples, and the small multiples then being placed against the wall. The second set shows a different 3D barchart being pushed into a virtual wall, and the bars stack themselves on top of each other into a 2D stacked barchart. The right part of the image shows a state machine of this morph. The start node connects to a 3D barchart state node, which is connected to two transition nodes: partitioning and stacking. The partitioning node is connected to a faceted barchart state node, and the stacking node is connected to a stacked barchart state node. Edges connecting these show signals describing conditions for their traversal. The three state nodes are connected to the exit node of the state machine.}
  \label{deimos_fig:teaser}
\end{teaserfigure}

\settopmatter{printfolios=true}
\maketitle

\section{Introduction} \label{deimos_sec:introduction}
Immersive environments, such as virtual and augmented reality (VR/AR), offer people a platform for human-computer interaction that utilises a variety of human senses and a range of physical human interactions. 
Compared to traditional desktop interaction, immersive environments offer users a more natural and \textit{embodied} experience of interaction \cite{dourishWhereActionFoundations2001}. 
Affordances for interaction can be embedded directly within virtual objects, allowing people to use their bodies to physically act upon those objects in a manner that leverages proprioception \cite{mineMovingObjectsSpace1997}. 
In the same way real-world objects can morph and change shape in response to physical actions, so should embodied representations of data in Immersive Analytics \cite{marriottImmersiveAnalytics2018}.
Interaction is crucial in data visualisation to handle complexity and allow changes to views \cite{munznerVisualizationAnalysisDesign2014}.
When an embodied visualisation is acted on by a user, it may undergo a transition in its visual state reflecting a change in encoding of data to representation. Animation is a very common technique to help users naturally keep track of such visual changes in statistical graphics \cite{heerAnimatedTransitionsStatistical2007,robertsonAnimatedVisualizationMultiple2002}.

Animation that preserves congruency between changes in data and its visual representation \cite{tverskyAnimationCanIt2002} has been demonstrated to confer benefits in myriad situations. It can aid decision-making in certain tasks  \cite{gonzalezDoesAnimationUser1996}, increase viewer engagement in data-driven stories \cite{heerAnimatedTransitionsStatistical2007, aminiHookedDataVideos2018}, and promote literacy of unfamiliar and/or complex visualisation designs \cite{ruchikachornLearningVisualizationsAnalogy2015,wangNarvisAuthoringNarrative2019}. However, these past explorations of animation in visualisation do not consider deep integration of animation and user interaction \cite{zongAnimatedVegaLiteUnifying2022}. 
Since embodied interaction relies on gestural congruency between the interaction and resulting visual changes, interaction and animation both clearly go hand in hand for embodied Immersive Analytics applications.

However, compared to the decades of research and development of desktop-based data visualisation packages for animation (e.g.\ \cite{kimGeminiGeneratingKeyframeOriented2021,kimGeminiGeneratingKeyframeOriented2021,zongAnimatedVegaLiteUnifying2022,geCanisHighLevel2020,geCASTAuthoringDataDriven2021,thompsonDataAnimatorAuthoring2021}) and interaction (e.g.\ \cite{satyanarayanVegaLiteGrammarInteractive2017,bostockDataDrivenDocuments2011,zongAnimatedVegaLiteUnifying2022}), equivalent tools for Immersive Analytics lag far behind. While some Immersive Analytics research has investigated the combination of animation and interaction \cite{yangTiltMapInteractive2020,leeDesignSpaceData2022}, no work has yet presented a unified language and grammar for the definition of such immersive interactive animations. Moreover, despite the numerous toolkits supporting the authoring of immersive visualisations (e.g.\ \cite{sicatDXRToolkitBuilding2019,cordeilIATKImmersiveAnalytics2019,reipschlagerPersonalAugmentedReality2021,butcherVRIAWebBasedFramework2021}), none allow for the rapid design and prototyping of embodied interactions: a glaring gap in the literature given the prevalence of embodiment in Immersive Analytics \cite{buschelInteractionImmersiveAnalytics2018}.

Therefore in this paper we introduce Deimos: a declarative grammar for authoring \textbf{d}ynamic \textbf{e}mbodied \textbf{i}mmersive \textbf{mo}rph\textbf{s} for immersive visualisations. We use the term morph to signify an embodied visualisation's ability to change shape when actions are performed on it by a user.
In contrast to traditional animated transitions, morphs are \textit{adaptive} and can be applied to any data visualisation in the environment that matches the partial visualisation specification of one of the morph's \textit{states}. \textit{Transitions} connect these states through animation that can be controlled by \textit{signals}: data streams which stem from \textit{embodied} user interaction. These are specified using the Deimos grammar and are written in JSON. The adaptivity of morphs allows them to be used in both analysis and presentation, depending on the degree of specificity of the morph.

We begin by detailing a set of design goals that allow morphs to leverage the strengths of immersive environments not present on desktops (Section \ref{deimos_sec:design-goals}). We then introduce the Deimos grammar itself, detailing its components, primitives, and specification (Section \ref{deimos_sec:grammar}). Next, we describe a prototype implementation of the Deimos grammar (Section \ref{deimos_sec:prototype}), developed in Unity as an extension to the DXR toolkit by Sicat et al.\ \cite{sicatDXRToolkitBuilding2019}. To demonstrate the expressivity of Deimos, we present an example gallery of morphs created in Deimos which highlights key characteristics of the grammar (Section \ref{deimos_sec:example-gallery}). We also conducted a user study in which six Immersive Analytics researchers used Deimos to create their own morphs. Through semi-structured interviews with these participants, we gauge the usability of Deimos (Section~\ref{deimos_sec:user-study}) and elicit discussion topics and future research directions for morphs (Section~\ref{deimos_sec:discussion}).

Our contributions include both engineering efforts and theoretical knowledge, and are summarised as follows:
\begin{enumerate}
    \item A grammar for the declaration of dynamic, embodied, interactive animated morphs in immersive environments called Deimos, and an implementation of the grammar in Unity.
    \item An example gallery of interactive morphs, and a user study and semi-structured interview with six Immersive Analytics researchers that validates the design, implementation, and usability of the Deimos grammar.
    \item An open-source toolkit that enables rapid design and prototyping of embodied interactions for Immersive Analytics which can accelerate future research in this area.
    \item A conceptualisation of how morphs can be defined as keyframe animations but be later applied as presets \& templates during analysis and/or presentation in VR/AR.
    \item A shift towards animation that is designed around and driven by (embodied) interaction, as opposed to existing methods that are mostly driven by the data.
\end{enumerate}

\section{Related Work} \label{deimos_sec:related-work}
\subsection{Interactive Animated Transitions on 2D Screens}
When a visualisation changes between visual states, animation is commonly used to help viewers maintain awareness of how data marks have changed throughout the transition \cite{heerAnimatedTransitionsStatistical2007,robertsonAnimatedVisualizationMultiple2002}, thus minimising change blindness \cite{munznerVisualizationAnalysisDesign2014}.
Various grammars and toolkits have been developed to aid designers in creating animated 2D statistical graphics for use in data-driven storytelling, such as Gemini~\cite{kimGeminiGrammarRecommender2021} and Gemini{\textsuperscript{2}}~\cite{kimGeminiGeneratingKeyframeOriented2021}, Canis~\cite{geCanisHighLevel2020} and CAST~\cite{geCASTAuthoringDataDriven2021}, and DataAnimator~\cite{thompsonDataAnimatorAuthoring2021}. These all fundamentally use keyframe animation, which has been shown to be the preferred paradigm of animation designers \cite{thompsonUnderstandingDesignSpace2020}.
Earlier work by Tversky et al.\ \cite{tverskyAnimationCanIt2002} however could not find strong evidence of animated graphics being superior to static ones, especially as animations were often too complex or fast to be accurately perceived. They instead suggested that interactivity may be one way to capitalise on the strengths of animation by allowing users to directly control its playback (start, stop, rewind, etc.). Indeed, later research found that combining interactivity with animations can improve outcomes for certain data analysis tasks (e.g.\ \cite{robertsonEffectivenessAnimationTrend2008,abukhodairDoesInteractiveAnimation2013}).
More recent work by Zong and Pollock et al.\ \cite{zongAnimatedVegaLiteUnifying2022} formalised interactive animation in the form of Animated Vega-Lite, an extension to Vega-Lite~\cite{satyanarayanVegaLiteGrammarInteractive2017} which adds a time encoding channel and event streams to enable interactive animations for use in data analysis. Such interactive animations (e.g.\ \cite{robertsonEffectivenessAnimationTrend2008,abukhodairDoesInteractiveAnimation2013,zongAnimatedVegaLiteUnifying2022,roslingBestStatsYou2007}) oftentimes expose their animation controls via a time slider and toggleable start/stop button.
A good example of more direct interaction with conventional 2D animations is that of DimpVis by Kondo and Collins \cite{kondoDimpVisExploringTimevarying2014}. Through direct manipulation, users can touch a mark to select it, revealing a ``hint path'' that they can drag their finger along. This causes the visualisation to temporally navigate forwards or backwards using animation, with the selected mark following the hint path. The subsequent work on Glidgets by Kondo et al.\ \cite{kondoGlidgetsInteractiveGlyphs2014} followed a similar premise but for dynamic graphs.

Of course, our work is differentiated from that of previous works by its immersive nature. We introduce new concepts and ideas to accommodate the shift to immersive environments, as we later detail in Section~\ref{deimos_sec:design-goals}.

\subsection{Embodied Interaction and Metaphors for Immersive Animations} \label{deimos_ssc:embodied-interaction-and-metaphors}
Immersive Analytics is characterised by the use of interactive, engaging, and embodied analysis tools \cite{marriottImmersiveAnalytics2018}. As such, there is a desire to move away from WIMP-based controls in favour of more direct, embodied styles of interaction \cite{cordeilImAxesImmersiveAxes2017,buschelInteractionImmersiveAnalytics2018}. 
In embodied interaction \cite{dourishWhereActionFoundations2001}, affordances are embedded within the artefact (in our case the data visualisation) itself, re-framing computational processes and operations as direct interactions of one's body with the physical world \cite{williamsInteractionParticipationConfiguring2005,fishkinEmbodiedUserInterfaces2000}. This approach, as Dourish \cite{dourishWhereActionFoundations2001} notes, moves the user interface into the background where it is no longer the centre of attention.
Embodied interaction is capable of leveraging metaphors \cite{lakoffMetaphorsWeLive2008}, which can make it easier to remember interaction techniques and help users develop their mental model of the target domain \cite{carrollChapterInterfaceMetaphors1988}. Such metaphors have been extensively used in embodied Immersive Analytics research as a result, typically involving mid-air input. ImAxes by Cordeil et al.\ \cite{cordeilImAxesImmersiveAxes2017} used several interaction metaphors, such as direct manipulation to compose visualisations based on the proximity and relative orientation of embodied axes (a similar metaphor was also employed using the MADE-Axis by Smiley et al.\ \cite{smileyMADEAxisModularActuated2021}), and a ``throw away'' metaphor to delete these visualisations. FIESTA by Lee et al.\ \cite{leeSharedSurfacesSpaces2021} used a similar throwing metaphor but for pinning visualisations onto surfaces in the environment. FiberClay by Hurter et al.\ \cite{hurterFiberClaySculptingThree2019} used a ``grab'' metaphor for translating, rotating, and scaling a 3D trajectory visualisation.

Embodied interaction has also been used to directly control immersive animated transitions. Tilt Map by Yang et al.\ \cite{yangTiltMapInteractive2020} is a visualisation that transforms between three states: a choropleth map, prism map, and barchart. As the visualisation is tilted using a VR controller, the visualisation is interpolated between the three states based on the tilt angle.
More interesting is the recent work by Lee et al.\ \cite{leeDesignSpaceData2022} which demonstrated the use of the visualisation's spatial context as part of the metaphor. They described techniques for transforming visualisations between 2D and 3D, such as ``extruding'' a 2D visualisation into 3D using a ``pinch and pull'' gesture. For the technique to be valid however, the 2D visualisation must also be placed against a physical 2D surface. Through this, the metaphor is not only of the visualisation being extruded, but also of it being taken from a surface and ``brought out into'' space. Both of these works \cite{yangTiltMapInteractive2020,leeDesignSpaceData2022} also demonstrate a high level of gestural congruency between the interaction and the visualisation that is manipulated, which is vital in embodied interaction \cite{johnson-glenbergEmbodiedScienceMixed2017,johnson-glenbergImmersiveVREducation2018}. For example, the aforementioned extrusion technique described by Lee et al.\ \cite{leeDesignSpaceData2022} causes the visualisation to expand at the same rate as the hand is being pulled, directly mapping the extent of the extrusion to the user's hand position.

While other works do use animations in prototype implementations (e.g.\ \cite{hayatpurDataHopSpatialData2020,flowimmersiveinc.DataStorytellingImmersive2022,cordeilIATKImmersiveAnalytics2019}), animation has largely been used to maintain awareness during transitions and has not been the focal point of the research (unlike that of Yang et al.\ \cite{yangTiltMapInteractive2020} and Lee et al.\ \cite{leeDesignSpaceData2022}). Therefore in this work we further explore the use of embodied interaction to control visualisation animations in immersive environments.

\subsection{Toolkits and Grammars for Immersive Analytics}
In recent years, many toolkits and frameworks have emerged to support research and development in Immersive Analytics. Some specialised toolkits have been developed which focus on specific application cases. MIRIA~\cite{buschelMIRIAMixedReality2021} allows user experiment data such as head and hand movements to be replayed in an AR environment for in-situ analytics.
RagRug~\cite{fleckRagRugToolkitSituated2022} is a situated analytics toolkit that updates immersive visualisations in either VR or AR through the use of a distributed data flow from the Internet of Things and NODE-Red.

Toolkits have also been developed to facilitate more generic visualisation authoring in immersive environments. While certainly not as mature as desktop-based packages such as gg2plot \cite{wickhamLayeredGrammarGraphics2010} and D3 \cite{bostockDataDrivenDocuments2011}, they typically provide a strong foundation that can and have been extended in subsequent works. These toolkits can largely be distinguished by how visualisations are created by the user. IATK~\cite{cordeilIATKImmersiveAnalytics2019} and u2vis~\cite{reipschlagerDesignARImmersive3DModeling2019} primarily expose their authoring tools through a GUI---typically through the Inspector window of the Unity game engine's editor. In contrast, DXR~\cite{sicatDXRToolkitBuilding2019} and VRIA~\cite{butcherVRIAWebBasedFramework2021} facilitate visualisation authoring using human-readable JSON files. A grammar defines the syntactical rules of this JSON file, which is then interpreted by the system to produce the visualisation. In the case of both DXR and VRIA, the grammar is based on Vega-Lite's grammar \cite{satyanarayanVegaLiteGrammarInteractive2017}. Declarative grammars such as these have proven to be popular in data visualisation (e.g.\ \cite{satyanarayanVegaLiteGrammarInteractive2017,zongAnimatedVegaLiteUnifying2022,kimGeminiGrammarRecommender2021,geCanisHighLevel2020}) as they separate how a visualisation is defined from how it is created by the system. These declarative grammars can also make it easier to author data visualisations, thus leading to more rapid prototyping of ideas.

A common limitation in Immersive Analytics toolkits however is their support for interactivity. While toolkits like IATK~\cite{cordeilIATKImmersiveAnalytics2019} and DXR~\cite{sicatDXRToolkitBuilding2019} provide built-in methods for interacting with the visualisation such as brushing and range filtering, they do not expose user-friendly means to create new interactions and instead require extending the source code itself. In contrast, our work aims to devise a grammar that can enable \textit{interactive} animated transitions in immersive environments. As a result, our work contributes a grammar that can support both authoring of immersive animated transitions and help design new (embodied) interaction techniques.
\section{Deimos Design Goals} \label{deimos_sec:design-goals}
The shift from 2D to 3D is more than just a third spatial encoding.
Early in the development of Deimos, we identified several key differences between animated transitions in immersive and non-immersive environments that give rise to new research challenges.
These challenges were rephrased and synthesised into three design goals (DG) which influenced the creation of the Deimos grammar, allowing us to focus on the novel characteristics of immersive headsets and environments, in turn opening up further design opportunities. Section~\ref{deimos_sec:grammar} will explain the grammar itself and highlight how it addresses these design goals.

\subsection{DG1: Morphs should be adaptable and flexible} \label{deimos_ssc:design-goal-1}
Most animated transition grammars allow for rapid prototyping between the specification and the resulting animation. A low viscosity authoring process is particularly important when creating interactive animations for data analysis~\cite{zongAnimatedVegaLiteUnifying2022}, allowing for fast and easy changes in the specification. The ability to rapidly prototype is facilitated by the constant access of keyboards for text input and pointing devices (i.e.\ mice) in desktop environments. In contrast, a challenge of immersive environments is that they often lack a convenient and comfortable form of text input that is required to write textual specifications, especially in VR or in highly mobile AR contexts. While a GUI can help facilitate this authoring process in VR/AR, designing a GUI is premature if there is no underlying grammar to support it, especially in such a novel environment.

To resolve this conflict, we take an approach inspired by Lee et al.'s recent work~\cite{leeDesignSpaceData2022}. Many animated transition grammars treat transitions as a bespoke set of changes applied to visualisations predefined by the animation designer. Instead, we treat animated transitions as discrete operations that analysts can use to apply changes to their visualisations during their analysis. For example, the analyst might apply an animated transition that adds another spatial encoding to their visualisation, or converts a 3D barchart into a faceted 2D barchart. This turns animated transitions into a catalogue of adaptive and flexible operations that can be applied to immersive visualisations by analysts depending on the situation and goals.
In this way, there exists two types of users of Deimos: immersive analytics system designers who use the grammar to create a catalogue of animated transitions in a desktop environment (e.g.\ Unity editor), and data analysts in VR/AR who use said animated transitions in their workflows and either do not have access to or are unfamiliar with the grammar.
This necessitates a functional shift in grammar design, moving from highly tailored transitions with known data fields and encodings to generic transitions that operate on baseline idioms. As a result, any given transition specification can be reused across multiple visualisations, so long as they meet the baseline criteria specified by the author (e.g.\ be a barchart, have no \textit{z} encoding).

\subsection{DG2: Morphs should support embodied interaction} \label{deimos_ssc:design-goal-2}
Animated transition grammars (e.g.\ \cite{kimGeminiGrammarRecommender2021, geCanisHighLevel2020, thompsonDataAnimatorAuthoring2021}) have paid little attention to how transitions are triggered and controlled. In cases where these grammars do (e.g.\ \cite{zongAnimatedVegaLiteUnifying2022}) it is limited to WIMP-style controls, with practitioners using similar input methods for their narrative visualisations (e.g.\ play button~\cite{roslingBestStatsYou2007}, linear slider/scroll~\cite{yeeVisualIntroductionMachine2015}).
In contrast, immersive environments rely on a completely different interaction paradigm which goes beyond the desktop and is both embodied (e.g.\ \cite{hurterFiberClaySculptingThree2019, cordeilIATKImmersiveAnalytics2019}) and spatial in nature (e.g.\ \cite{hubenschmidSTREAMExploringCombination2021, buschelInvestigatingUseSpatial2017}).
Novel language primitives are needed to support embodied interaction as existing ones (i.e.\ streams in Animated Vega-Lite \cite{zongAnimatedVegaLiteUnifying2022}) do not adequately express relationships between entities, especially desktop-based grammars. One such relationship is that of the user and the visualisation itself: which part of the user is performing the interaction (e.g.\ hand, head), and which part of the visualisation contains the affordance to be interacted with (e.g.\ mark, axis).
Spatial relationships and interaction also play a significant role in immersive environments \cite{buschelInvestigatingUseSpatial2017,langnerMARVISCombiningMobile2021,hubenschmidSTREAMExploringCombination2021}---which is not generally the case in non-immersive environments. For example, an immersive transition may be controlled based on the position of a handheld relative to a table~\cite{buschelInvestigatingUseSpatial2017}. By supporting this, immersive transitions become spatially aware. There can also be a relationship between the visualisation and its immediate environment, allowing immersive transitions to become context-aware \cite{svanaesContextAwareTechnologyPhenomenological2001,deyConceptualFrameworkToolkit2001}. An example of this is the aforementioned ``extrusion'' techniques by Lee et al.\ \cite{leeSharedSurfacesSpaces2021} which require the 2D visualisation to be on a surface to be usable.

By expanding the Deimos grammar to support this paradigm, we enable a richer design space of visualisation transitions not otherwise possible on desktop environments, as they allow users to ``reach through'' and interact with their data in a more embodied and engaging manner~\cite{dourishWhereActionFoundations2001}.
It should be noted however that the actual design of such embodied interactions is left up to the end-users of Deimos. We decide not to enforce best practices in the grammar, such as the use of easy to understand metaphors \cite{lakoffMetaphorsWeLive2008,carrollChapterInterfaceMetaphors1988} and proper gestural congruency \cite{johnson-glenbergEmbodiedScienceMixed2017,johnson-glenbergImmersiveVREducation2018}. Instead, we ensure Deimos is designed to allow said best practices to be followed---much in the same way that conventional programming languages do not enforce best practices.

\subsection{DG3: Morphs should still support conventional approaches}
While the two previous design goals are intentionally forward-thinking, we still want Deimos to be rooted in the same foundational elements as existing grammars. This is to both ensure that Deimos follows tried and true concepts and theories, and also to preserve a sense of familiarity for users of the grammar---especially for those new to immersive analytics. This includes the use of keyframe animation as the chief animation paradigm~\cite{thompsonDataAnimatorAuthoring2021}, the ability to specify timing and staging rules to customise the animation, and supporting WIMP-based interaction in hybrid immersive analytics setups or via immersive UX elements (e.g.\ \cite{microsoftMixedRealityUX2021}).
Moreover, while DG1 advocates for generalised transitions that can be applied to a wide range of visualisations, Deimos should still allow for highly customised transitions that affect predefined visualisations created by designers. This is to allow animated transitions in Deimos to still be useful in controlled situations such as immersive data-driven storytelling.
Therefore, our grammar should support both ends of two orthogonal spectrums: support both WIMP and embodied interaction to control and interact with animated transitions; and support animated transitions that are either highly generalised and can apply to any visualisation, or highly specific and apply only to a particular visualisation in a controlled context.
\section{The Deimos Grammar} \label{deimos_sec:grammar}
Deimos is a declarative grammar used to specify \textit{transitions} between \textit{states} (keyframes), as well as the \textit{signals} (interactions) used to control them. The grammar is largely based on the design goals listed in Section~\ref{deimos_sec:design-goals} and prior work by Lee et al.\ \cite{leeDesignSpaceData2022} on visualisation transformations.
The Deimos grammar was developed in conjunction with its toolkit implementation (Section~\ref{deimos_sec:prototype}) through an iterative process. At each iteration, a working version of the grammar was defined and the toolkit was updated to support it. We created new example morphs at each iteration to test the new features added to the grammar, and maintained prior examples to validate any adjustments to the grammar (similar to unit testing). Many of these examples can be seen in Section~\ref{deimos_sec:example-gallery}. We continued this process until we felt that the grammar sufficiently met our design goals.
The target audience of the grammar are developers and designers of immersive analytics systems. The morphs they create are then used by analysts in VR/AR.

A Deimos specification can formally be described as a three-tuple (elements suffixed with ``?'' are optional):

\begin{center}
\textit{Morph := (states, signals?, transitions)}
\end{center}

\begin{figure}[tb]
    \centering
    \includegraphics[width=\linewidth]{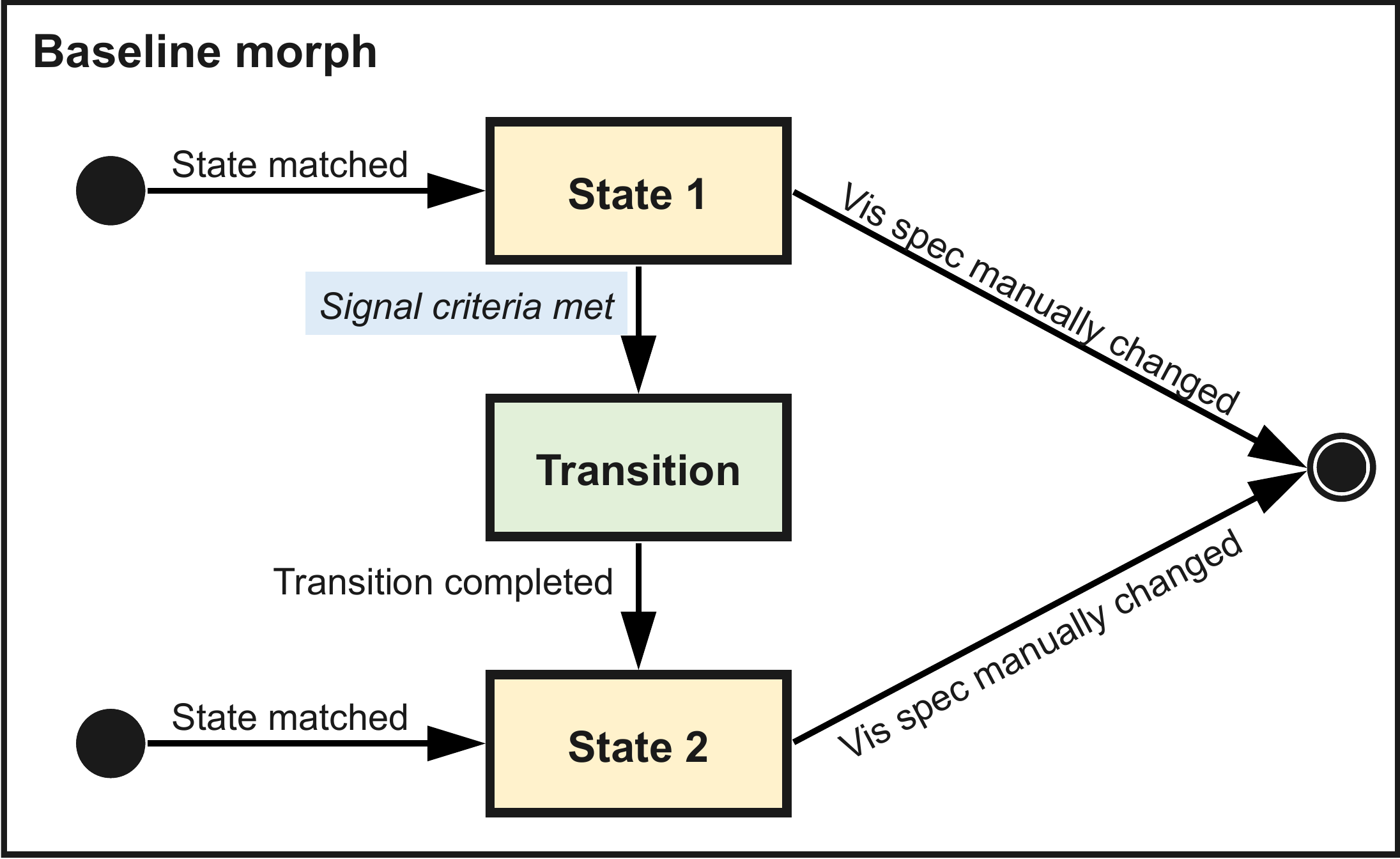}
    \caption{Baseline state machine for Deimos morphs showing a single unidirectional transition. More states and transitions can be added to the state machine with their own signal criteria, with support for bidirectional transitions.}
    \Description{A state machine for a baseline morph. A node labelled ``State 1'' is connected with a directed edge to a node labelled ``Transition'', and is labelled with ``Signal criteria met''. The latter node is connected with another directed edge to another node labelled ``State 2'', and is labelled with ``Transition completed''. Two entry nodes connect to the two State nodes with labels reading ``State matched''. The two State nodes are also connected to an exit node reading ``Vis spec manually changed''.}
    \label{deimos_fig:deimos-state-machine-1}
\end{figure}

These components constitute what we call a \textit{Morph}, the term signifying an embodied visualisation's ability to dynamically change shape and morph from one state to another via transitions upon matching certain conditions.
A morph can be modelled as a state machine (Figure~\ref{deimos_fig:deimos-state-machine-1}). A visualisation in the immersive environment only enters a morph's state machine when it matches one of its \textit{states}. The state node that was matched with determines the possible \textit{transition} nodes that it can access. These transition nodes are where changes are actually made to the visualisation, and are only entered when specified criteria are met. These criteria take the form of \textit{signals}, which are streams of data typically generated by user interaction. They can also be used to control the behaviour of transitions themselves.

Morphs are an extension to any immersive visualisation authoring system already in place. That is, visualisations can still be manipulated in their usual way, but can have morphs applied to them should the relevant conditions be met. In this way, morphs serve purely to augment existing authoring techniques rather than supplanting them outright. When a visualisation is modified by the user in a manner external to the morph, it exits the morph state machine. It may then immediately re-enter following the same rules as before. A visualisation can have multiple morphs (and therefore state machines) active simultaneously. Multiple morphs can also be applied to the same visualisation concurrently, so long as the properties and encodings they affect do not overlap. The same morph specification can also be active across multiple eligible visualisations. This ability for the state machine to adapt to different visualisation configurations through a set of rules and conditions is what helps it satisfy DG1.

\begin{figure}[tb]
    \centering
    \includegraphics[width=\linewidth]{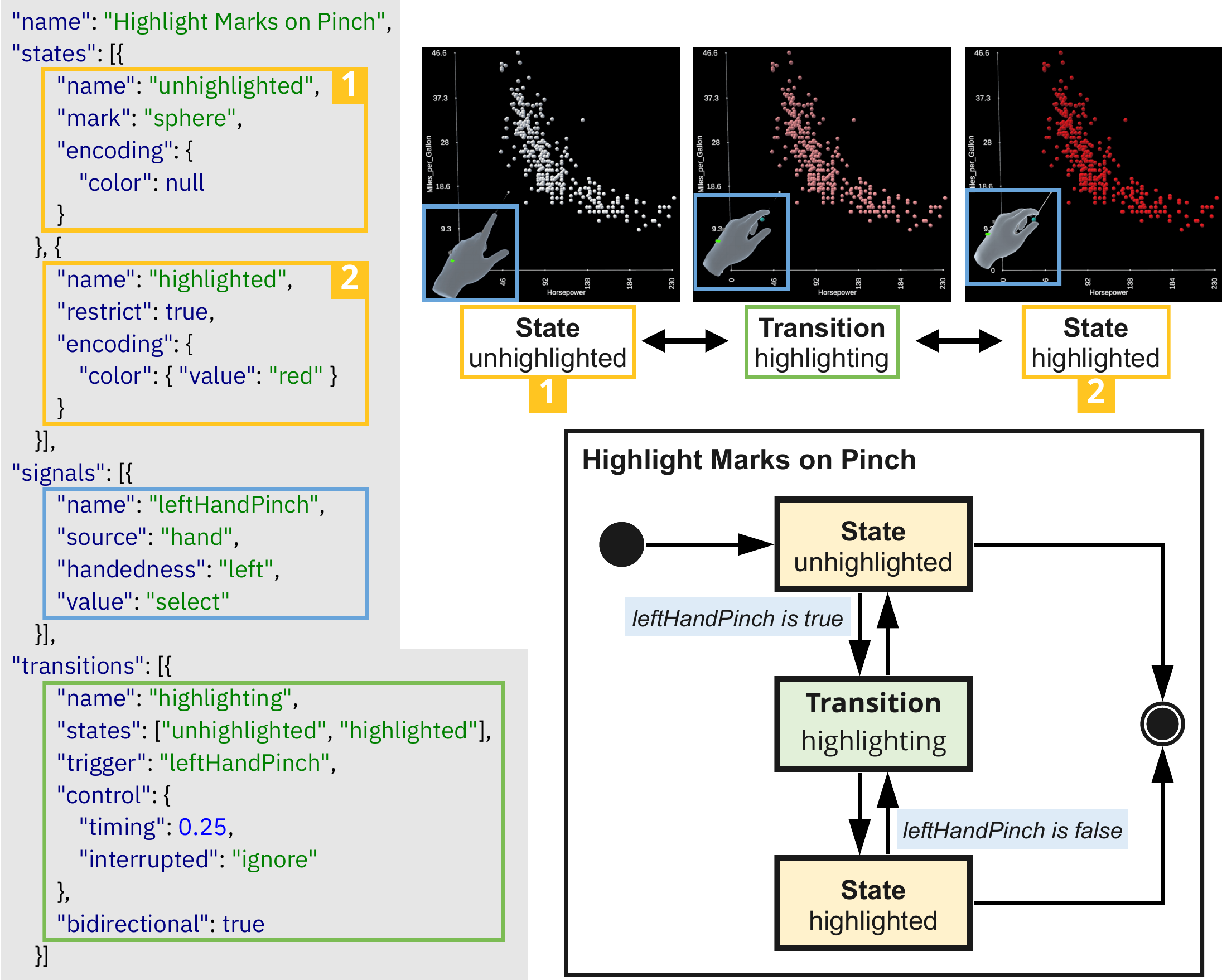}
    \caption{A basic example of a morph changes the mark colour of uncoloured visualisations to red whenever the left hand performs a pinch gesture. Colour-coded boxes denote the same component in different representations. Left: The morph specification. Top right: Still images of this morph being applied to a 2D scatterplot in an immersive environment. Bottom right: The state machine for this morph. The \texttt{``restrict'': true} (shown in the left-hand box labelled with ``2'') prevents the morph from starting at the \textit{highlighted} state, and \texttt{``bidirectional'': true} (shown at the end of morph specification) allows the transition to function in both directions.}
    \Description{An image composed of three parts. The first part shows a morph specification written in JSON. The second part shows a series of still images in which a hand does a pinch gesture, and causes the marks on a 2D scatterplot to turn red. The second part shows the state machine of this morph. The following three nodes are connected in a linear order: a ``unhighlighted'' state node, a ``highlighting'' transition node, and a ``highlighted'' state node.}
    \label{deimos_fig:deimos-json}
\end{figure}

Morph specifications are written and stored as standalone JSON files. The use of JSON is very common amongst related grammars and allows for the separation between grammar and implementation (i.e.\ portability). A JSON schema provides auto-completion and tooltips for writing morph specifications with supported text editors. Figure~\ref{deimos_fig:deimos-json} shows a basic example of a morph specification, and how it translates to the immersive environment and the state machine. The three main components of morphs are annotated with coloured boxes: states in yellow, signals in blue, and transitions in green. The same colour coding is used across all other figures.
The rest of this section will explain in general terms what these components do.

\subsection{States} \label{deimos_ssc:states}
A morph is comprised of at least two state specifications. A state can be defined by the following tuple:

\begin{center}
\textit{state := (name, restrict?, partial visualisation specification)}
\end{center}

The \textit{name} property is a unique case-sensitive string used to reference this state specification in a transition Section~\ref{deimos_ssc:transitions}).
The \textit{restrict} property is a Boolean that if set to \texttt{true} will remove the entry point associated with the state's node on the state machine Figure~\ref{deimos_fig:deimos-json} for an example). This prevents a morph from starting at that state, making it only accessible via interconnecting transition(s). This is useful if it is illogical for a morph to start at that state, such as in unidirectional transitions.
\textit{Partial visualisation specification} is an arbitrary number of properties and components in the state object that all follow the same declarative notation as an actual visualisation. In other words, its syntax is the same as the visualisation package used in the system. For our implementation of Deimos, this is the DXR grammar~\cite{sicatDXRToolkitBuilding2019} which in turn is based on the Vega-Lite grammar~\cite{satyanarayanVegaLiteGrammarInteractive2017}. In the context of the DXR grammar, a partial specification can consist of any number of view-level properties (e.g.\ \textit{mark}, \textit{depth}) and/or encoding-level properties declared inside of an \textit{encoding} component (e.g.\ \textit{x}, \textit{color}).
The partial specification serves two purposes: (i) to determine if a visualisation matches (and therefore enters) this state; and (ii) to generate the keyframe used in the transition.

\subsubsection{State matching process} \label{deimos_sss:state-matching-process}
Any visualisation properties specified as part of the \textit{partial visualisation specification} in a state are used in the matching process against active visualisations. It is important to differentiate between the two types of specifications being used in this process: the visualisation specification created by the end-user, and the state specification (i.e.\ the \textit{partial visualisation specification}) that exists as a part of the state component in a morph.
Generally speaking, for a state specification to be matched against a visualisation specification, all properties defined in the former should also be defined in the latter, including their associated values.  For example, if the state has \texttt{``color'': \string{``type'': ``quantitative''\string}}, then the visualisation must also have a \texttt{color} encoding with the same \texttt{type} for it to match.
As a rule of thumb, the fewer properties defined in the state specification, the more likely a visualisation can match successfully and have morphs applied to it. The opposite is also true, with more properties in the state specification making it less likely for any visualisation to match successfully. This effectively forms a spectrum. Morphs can be highly generic and can apply to many visualisations, allowing for adaptive morphs as per DG1. They can also only apply to specific datasets and field names, allowing for highly tailored morphs that are used in controlled environments as per DG3.

Deimos provides several primitives which affect the matching process that can be used in place of any JSON value in the state specification. They allow for more nuanced control over which visualisations can and cannot match, and are useful to prevent morphs from being accidentally applied to incompatible visualisations.
Note that this is not an exhaustive set of primitives. While they were adequate for the purposes of this work, the grammar can easily be extended to include more if need be.

\begin{itemize}
    \item \textbf{``*'' (wildcard)}: The property should be in the visualisation but its value can be anything.
    \item \textbf{An inequality expression}: The property should be in the visualisation and its value should satisfy the inequality. Only applicable to numeric properties. e.g.\ \texttt{``value'': ``>= 100''}.
    \item \texttt{null}: The property should not be included in the visualisation regardless of its value.
\end{itemize}

\subsubsection{Keyframe creation process} \label{deimos_sss:keyframe-creation-process}
When a visualisation matches a state and one of its connecting transitions is activated, keyframes are generated for both initial and final states. These keyframes are used for actual animation during the transition.
The initial keyframe is always the active visualisation's specification prior to the transition. No changes need to be made to it as it already matches the properties of the state itself.
The final keyframe is created by modifying the initial keyframe using the following set of rules: (i) visualisation properties that are defined in the initial state but not in the final state are removed; (ii) properties that are not defined in the initial state but are defined in the final state are added; and (iii) properties defined in both states are set to the final state's value.

As with the state machine process (Section~\ref{deimos_sss:state-matching-process}), Deimos provides primitives that can be used in place of any JSON value to refine the keyframe creation process. These primitives functionally act as placeholders which are later substituted with real values calculated at runtime, akin to the notion of variables. This allows for morphs to adapt to a wider range of situations without the need to hard-code field names, data types, etc. in morph specifications.
For the purposes of the state matching process, all of these primitives are treated as wildcards. Their values are resolved after the keyframes have been created but before the transition is applied. Once again, this is not an exhaustive list of primitives and can easily be extended if need be.

\begin{itemize}
    \item \textbf{JSON path accessor}: The value residing at the JSON path will be substituted into the property's value. Is either prefixed with ``this.'' to access a property from this keyframe, or ``other'' to access a property from the other keyframe which is being transitioned to/from. e.g.\ \texttt{``x'': ``this.\linebreak[0]encoding.\linebreak[0]y'', \texttt{``field'': ``this.\linebreak[0]encoding.\linebreak[0]size.\linebreak[0]field''}}.
    \item \textbf{A signal name}: The value emitted by the specified signal (Section~\ref{deimos_ssc:signals}) will be substituted into the property's value.
    \item \textbf{An expression}: The evaluated result of the expression will be substituted into the property's value. JSON path accessors and signal names can be used as variables. Only applicable to numeric properties. e.g.\ \texttt{``value'': ``other.\linebreak[0]encoding.\linebreak[0]size.\linebreak[0]value * 10''}.
\end{itemize}

All keyframes are stored throughout the entire lifespan of a morph. When the morph exits the state machine---the result of the associated visualisation having its specification manually changed by the user (Figure~\ref{deimos_fig:deimos-state-machine-1})---all stored keyframes are deleted.
Any added or changed properties will take their values from the state's keyframe if one already exists.
The main purpose for this is to handle situations where a property is removed by a transition in one direction, but needs to be added back in by a transition in the reverse direction. Without stored keyframes, the removed property would no longer be known and therefore could not be added back in.

\subsection{Signals} \label{deimos_ssc:signals}
In Deimos, a signal is the resulting value from a stream of data captured from input events, drawing inspiration from Vega's signals~\cite{satyanarayanDeclarativeInteractionDesign2014} and event-driven functional reactive programming principles~\cite{wanEventDrivenFRP2002}. Signals can be used in Deimos to: (i) be substituted as values in keyframes (Section~\ref{deimos_sss:keyframe-creation-process}); (ii) act as conditional triggers that control when a transition actually begins (Section~\ref{deimos_ssc:transitions}); and (iii) act as a tweening variable to control the progression of a transition (Section~\ref{deimos_ssc:transitions}). No type safety is enforced in Deimos. A morph may contain zero or more signal specifications.
Deimos has two main types of signals: signals that stem from some given source, and signals that evaluate a mathematical expression.

\begin{center}
    \textit{signal := sourceBasedSignal | expressionSignal}
\end{center}

\subsubsection{Source-based Signals} \label{deimos_sss:source-based-signals}
Source-based signals, as the name suggests, emit values from some input source. This is primarily from user interactions but could be extended to passively updating values from sensors, etc. We define two classes of source-based signals: deictic and non-deictic signals. Deictic signals express relationships between a source and target entity.
While they mainly serve to model direct manipulation which is commonly associated with embodied interaction (DG2), they can also model situations where there is no actual direct contact. Non-deictic signals capture everything else, although these are mainly input sources that do not require some target/context to make sense (e.g.\ mid-air hand gestures, input source states, sensor data). Their production rules are:

\begin{center}
\begin{tabular}{l}
    \textit{sourceBasedSignal := nonDeicticSignal | deicticSignal} \\
    \textit{nonDeicticSignal := (name, source, handedness?, value)} \\
    \textit{deicticSignal := (name, source, handedness?, target, criteria?, value)}
\end{tabular}
\end{center}

Both signal classes share the same three attributes.
The \textit{name} property references this signal in either a state (Section~\ref{deimos_sss:keyframe-creation-process}), an expression signal (Section~\ref{deimos_sss:expression-signals}), or a transition (Section~\ref{deimos_ssc:transitions}).
The \textit{source} property denotes the type of source that values are to be retrieved from (e.g.\ \texttt{hand}, \texttt{head}, \texttt{vis}, \texttt{ui}).
Certain sources can also specify the source's \textit{handedness} to distinguish between \texttt{left}, \texttt{right}, or defaulting to \texttt{any}.

For non-deictic signals, the \textit{value} property denotes what type of value to derive from the source, which is then emitted by the signal. This can either be the state of the user interaction (e.g.\ whether the hand is performing a \texttt{select} gesture) or the geometric properties of the source as an object in the immersive environment (e.g.\ \texttt{position} of the user's head). As previously mentioned, these are useful when some value of the input source is to be retrieved without it needing to be in the context of some other target or object. Figure~\ref{deimos_fig:deimos-json} shows an example of a non-deictic signal: it does not matter what the hand is touching so long as it is performing the pinch gesture.

Deictic signals model relationships between entities, and are based on the interaction section of the design space by Lee et al.\ \cite{leeDesignSpaceData2022}.
The \textit{target} property denotes the type of object that the source is attempting to target. This can either be a part of the visualisation (e.g.\ \texttt{mark}, \texttt{axis}), a separate object in the environment (e.g.\ \texttt{surface}), or part of the user themselves (e.g.\ \texttt{head}).
For the first two, a \textit{criteria} property needs to be included to determine the logic used in selecting the target (e.g.\ \texttt{select}, \texttt{touch}, \texttt{nearest}). This logic is needed when there are multiple potential target objects that could be selected.
Lastly, the \textit{value} property can be used to derive three types of values. First, it can derive values from the \textit{target} much in the same way as non-deictic signals do. For example, a \texttt{hand} source might target the \texttt{mark} that it is \texttt{select}ing, and the \texttt{position} of that mark is used as the value. Second, it can derive values from a comparison between the source and target. For example, a \texttt{vis} source might target the \texttt{surface} that it is \texttt{touch}ing, and the point of \texttt{intersection} between the vis and surface is used as the value. Third, a \texttt{boolean} value simply emits true if a target has been selected successfully, and false if no targets are selected.

Deictic signals in particular address the challenges in DG2 as they express relationships between entities, allowing morphs to react to direct interactions by the user (e.g.\ user's hand selects a mark). Of course, whether or not these interactions are truly embodied (i.e.\ it follows best practices) is dependent on how the morph designer uses deictic signals in conjunction with the grammar's other components.
Deictic signals also allow morphs to be spatially-aware \cite{buschelInvestigatingUseSpatial2017,langnerMARVISCombiningMobile2021,hubenschmidSTREAMExploringCombination2021}, as they can emit values that are based on spatial relationships between objects which can then be used to control the morph's behaviour (e.g.\ distance between user's head and the visualisation, orientation of two standalone tracked objects).
Lastly, deictic signals allow morphs to become context-aware \cite{svanaesContextAwareTechnologyPhenomenological2001,deyConceptualFrameworkToolkit2001}, as they can emit values derived from a visualisation's relationship with its environment (e.g.\ is the visualisation touching a surface, is the visualisation close to a particular object). This may then act as conditionals to allow/disallow the morph from triggering (Section~\ref{deimos_ssc:transitions}).

While not as critical to this work, the ability to facilitate WIMP-style interaction using these signals also helps fulfil DG3.

\subsubsection{Expression Signals} \label{deimos_sss:expression-signals}
\begin{center}
    \textit{expressionSignal := (name, expression)}
\end{center}

Expression signals allow for the arbitrary composition of signals using mathematical expressions. Their primary purpose is to modify and refine values emitted by source-based signals.
We choose to use expressions as they allow arbitrary calculations to be performed in a familiar manner, instead of designing a completely new and potentially confusing domain-specific language.
The \textit{name} property references this signal in the same way as source-based signals. The \textit{expression} property is a mathematical expression as a string. Basic mathematical operators can be used alongside select primitive functions (e.g.\ \texttt{normalise}, \texttt{distance}, \texttt{angle}).
As with all other primitives, the list of supported functions can easily be extended. Any type of signal can be used as a variable by referencing its name. As previously mentioned, no type safety is enforced, meaning the user has to be aware of the data types present in the expression.

Expression signals are similar to deictic signals in that they help further address the challenges in DG2, but are more powerful in comparison. For example, while deictic signals only allow for a single entity to be targeted, expression signals can combine two (or more) deictic signals together to calculate a new relationship between the targeted entities (e.g.\ distance between two marks selected by the user's hands).

\subsection{Transitions} \label{deimos_ssc:transitions}
A morph is comprised of at least one transition specification. They functionally connect two state specifications together in the state machine (Figure~\ref{deimos_fig:deimos-state-machine-1}). A transition can be defined by the following seven-tuple:

\begin{center}
\textit{transition := (name, states, trigger?, control?, bidirectional?, disablegrab?, priority?)}
\end{center}

The \textit{name} property serves to identify this transition especially when multiple transitions are involved.
The \textit{states} property is an array of two strings, corresponding to the names of the initial and final states in the transition respectively. Referencing states via their name in this manner helps with encapsulation, keeping all state related syntax separated from the transitions.
The \textit{trigger} property is an equality expression that activates the transition when it evaluates as true, but only when the visualisation matches the initial state in the \textit{states} property. The expression follows similar rules as expression signals (Section~\ref{deimos_sss:expression-signals}) but must return a Boolean value. Triggers are mainly used to let the user control when the transition is actually applied, usually as the result of some sort of input action or condition caused by the user. Not setting a trigger will cause the transition to be immediately applied when it enters the initial state.
The \textit{control} component is optionally used to further customise the behaviour of the transition. It is formally described by the following five-tuple:

\begin{center}
\textit{control := (timing?, easing?, interrupted?, completed?, staging?)}
\end{center}

The \textit{timing} property controls the duration of the transition. If a number is used, the transition will interpolate between the two state keyframes over the given duration in seconds. Alternatively, the name of a signal can be used, in which case the signal will be used as the tweening variable \textit{t}. This allows for the duration and direction of the interpolation to be controlled by the signal (and subsequently the user). In this situation, the transition will only begin when the signal is a value between 0 and 1, in addition to any other conditions. This defaults to 0 if not specified, which will result in jump cuts.
The \textit{easing} property applies an easing function to the transition, defaulting to a linear function if none is specified. Easing functions are commonly used in animations and help make animations look more natural. Functions that slow down the animation at the start and end can also make it easier to keep track of visual changes by making movement more predictable \cite{dragicevicTemporalDistortionAnimated2011}.
The \textit{interrupted} property determines what happens when the \textit{trigger} returns false whilst the transition is in progress. \texttt{initial} and \texttt{final} will cause the visualisation to immediately jump to the specified state. \texttt{ignore} will instead allow the transition to keep progressing until it naturally terminates. The \texttt{ignore} condition is particularly useful in cases where the \textit{trigger} may inadvertently return false mid-transition but the transition should still continue, acting as a sort of fail-safe. This defaults to \texttt{final}.
Similarly, the \textit{completed} property determines what happens when the visualisation naturally terminates, either remaining at the \texttt{final} state or resetting back to the \texttt{initial} state instantaneously. Using the \texttt{initial} condition may be useful if the transition should not cause any long-term changes to the visualisation, particularly if the animation is alone sufficient to serve its purpose~\cite{leeDesignSpaceData2022}. This also defaults to \texttt{final}.

The \textit{staging} property allows for specific visualisation properties to be staged. Name-value pairs can be specified where the name is the property to be staged, and the value is an array of two numbers between 0 and 1 that correspond to start and end percentages. The property will only be animated when the transition period is within the given range. Any property not specified will not be staged. Staging is a common feature of animated transition grammars~\cite{heerAnimatedTransitionsStatistical2007} and ours is no different. Note that the grammar does not support staggering.

The \textit{bidirectional} property of the transition, if set to true (default false), allows the transition to start and end in the reverse direction. All transition settings remain the same, except the \textit{trigger}, if specified, needs to return false in order for the reverse transition to activate. This serves mainly as a convenience function that prevents the need for two transition specifications to be written whenever a single bidirectional transition is desired. However, doing so is necessary in order to have distinct settings for either direction.
The \textit{disablegrab} property, if set to true (default false), will automatically disable the standard VR/AR grab action performed on the visualisation when the transition starts. This helps prevent visualisations from being inadvertently moved by the user when a transition's \textit{trigger} uses a similar grab gesture.
Lastly, the \textit{priority} property can be used to handle edge cases where multiple transitions due to similar \textit{trigger} conditions are activating on the same frame, but they conflict with the visualisation properties they modify. In this situation, the transition with the highest numbered priority will activate first, and all other conflicting transitions will be blocked. If priorities are equal, then the order in which they activate is random. The priority property defaults to 0.

\subsection{Satisfaction of Design Goals}
We now reiterate how our grammar satisfies the design goals listed in Section~\ref{deimos_sec:design-goals}.

For DG1, the use of partial visualisation states (Section~\ref{deimos_sss:state-matching-process}) and the keyframe creation process (Section~\ref{deimos_sss:keyframe-creation-process}) helps satisfy it. As the Deimos grammar is defined solely through JSON text, a library of generic morphs can be created in a development environment that has access to ergonomic text input (i.e.\ keyboards). When deployed in a production environment, the end-user in the immersive environment then has access to these (embodied) interactive morphs without needing to write the specifications themselves---a process which is notoriously difficult in VR and/or remote-AR environments. We provide a direct example of one such generic morph in Section~\ref{deimos_ssc:example-gallery-generic-specific}. Establishing this JSON-based grammar also sets the foundation for designing a GUI that is intended for use in VR/AR, much in the same way that CAST \cite{geCASTAuthoringDataDriven2021} is the GUI implementation of the Canis grammar \cite{geCanisHighLevel2020}. Through this, a morph author can rapidly prototype entirely in VR/AR.

For DG2, certain components such as deictic (Section~\ref{deimos_sss:source-based-signals}) and expression signals (Section~\ref{deimos_sss:expression-signals}) directly support embodied interaction, as these signals listen to user input and/or changes in the entities in the environment and thus the relationships between them. As previously stated in Section~\ref{deimos_ssc:design-goal-2}, the grammar intentionally does not enforce any best practices, including embodied interaction and animated transition principles. However, adherence to these guidelines is not isolated to any one component of a morph but instead across the entire specification. For example, even if direct manipulation is emulated through a deictic signal between the user's hands and the visualisation's marks, there would be little to no gestural congruency if the morph instead changed the visualisation's geometric size. Therefore, the ability of the grammar to express embodied interactions is dependent on the morph designer. We describe how a morph can use embodied interaction in a practical example in Section~\ref{deimos_ssc:example-gallery-embodied}. We also describe how morphs can be rapidly iterated on in order to test new (embodied) interaction ideas in additional examples in Section~\ref{deimos_ssc:example-gallery-prototyping}.

For DG3, certain source signals (Section~\ref{deimos_sss:source-based-signals}) allow for WIMP UI elements to be used to control morphs. This of course stands at odds with the embodied interactions of DG2, but our goal with Deimos is to support both ends of this theoretical spectrum. Section~\ref{deimos_sec:example-gallery} as a whole contains multiple examples of these more conventional types of morphs.
\begin{figure}[tb]
    \centering
    \includegraphics[width=\linewidth]{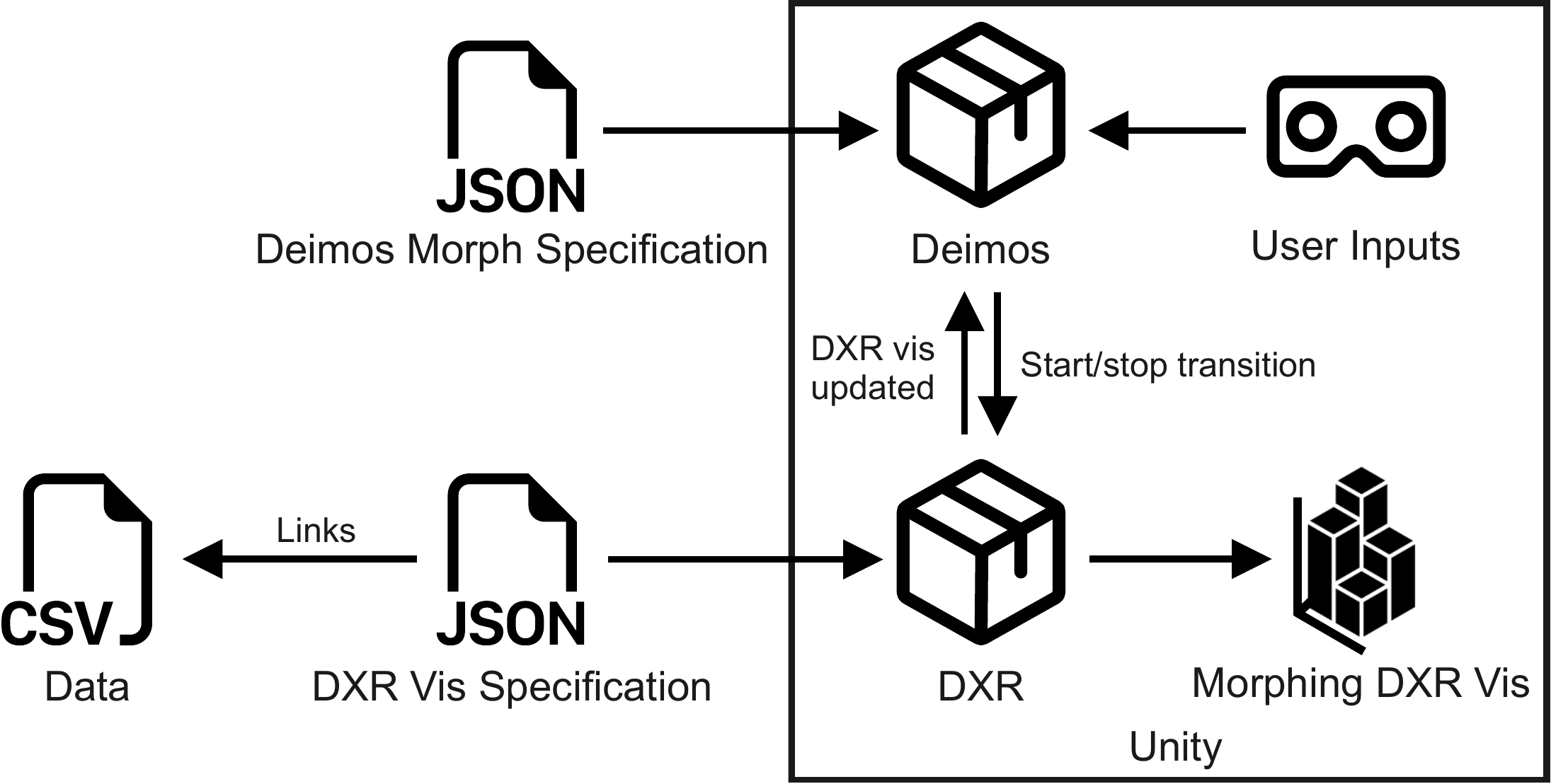}
    \caption{Overview of Deimos and how it interacts with our updated version of DXR \cite{sicatDXRToolkitBuilding2019}. A recreated version of the original DXR overview image is shown in the lower half.}
    \Description{A diagrammatic image Deimos. Deimos morph specifications connect to the Deimos package. User inputs also connect to the Deimos Package. Below it, a DXR vis specification connects to Data and the DXR package. The Deimos and DXR packages are connected to each other, with arrows indicating ``DXR vis updated'' and ``Start/stop transition''. The DXR package then connects to the Morphing DXR vis.}
    \label{deimos_fig:deimos-overview}
\end{figure}

\section{Deimos Implementation and Toolkit} \label{deimos_sec:prototype}
We created a prototype implementation of the Deimos grammar using the Unity game engine in order to demonstrate its concepts and use. Deimos is open source, with its source code and documentation available on a public GitHub repository\footnote{\url{https://github.com/benjaminchlee/Deimos}}.

\begin{figure}[tb]
    \centering
    \includegraphics[width=0.7\linewidth]{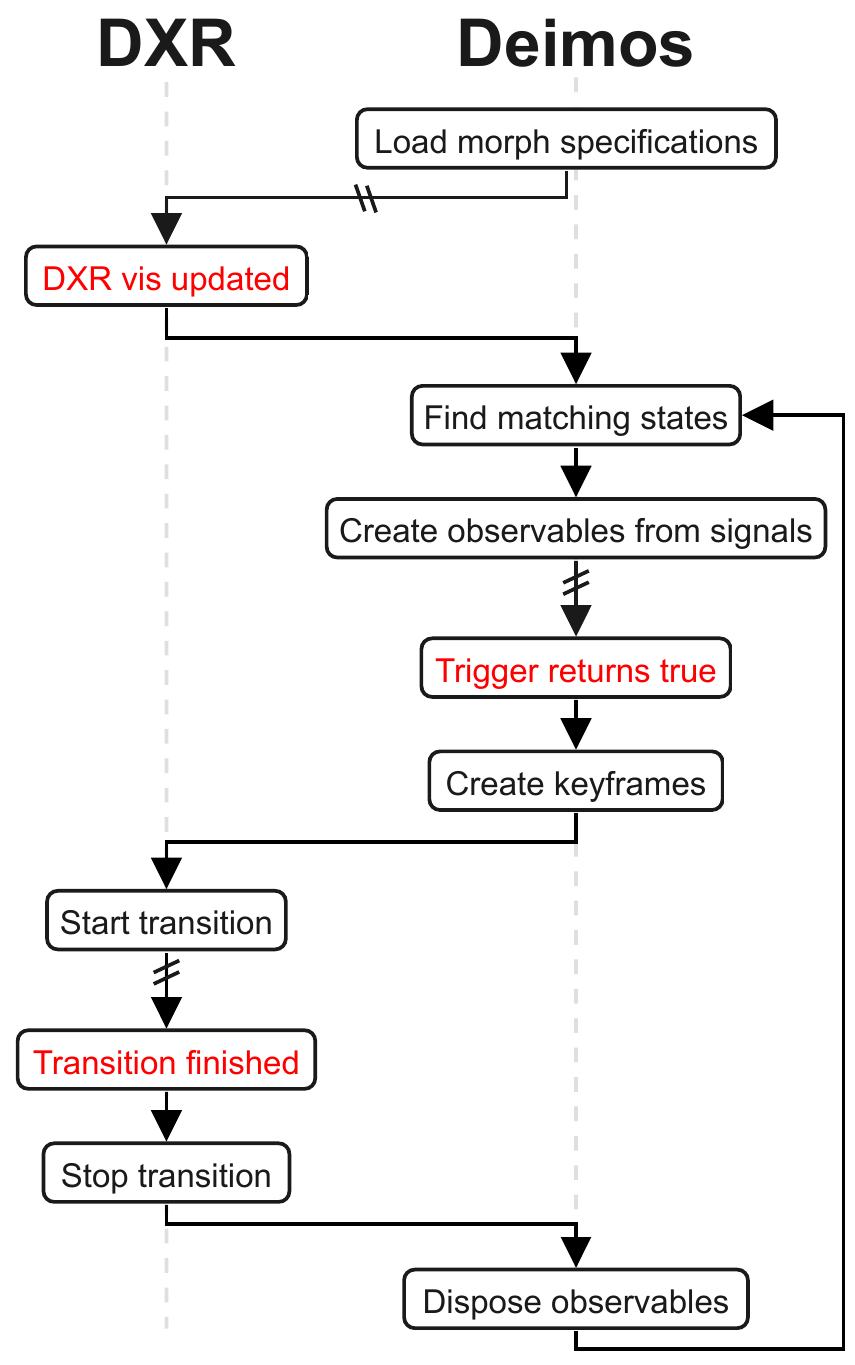}
    \caption{High-level overview of the Deimos pipeline and how it interacts with our updated version of DXR \cite{sicatDXRToolkitBuilding2019}. Red bubbles represent stages that wait for event(s) to fire before execution continues, also indicated by the preceding hatched arrow.}
    \Description{A linear flowchart of the Deimos pipeline. The nodes in linear order are Load morph specifications, DXR vis updated (in red), Find matching states, Create observables from signals, Trigger returns true (in red), Create keyframes, Start transition, Transition finished (in red), Stop transition, and Dispose observables. The last node is connected back to the Find matching states node.}
    \label{deimos_fig:deimos-pipeline}
\end{figure}

\subsection{Data Visualisations} \label{deimos_ssc:prototype-data-visualisations}
As Deimos is primarily an animated transition grammar, we need data visualisations to apply transitions to. 
We decided to use DXR by Sicat et al.\ \cite{sicatDXRToolkitBuilding2019} as the basis of our work. It is a toolkit developed for the Unity game engine designed for rapid prototyping of immersive data visualisations. The original DXR implementation provided support for an assortment of visualisation types, including scatterplots, barcharts, radial barcharts, and streamlines. These visualisations are specified in JSON files using an extended version of the Vega-Lite grammar \cite{satyanarayanVegaLiteGrammarInteractive2017}, adding support for the $z$ and \textit{depth} encodings. We use DXR instead of other toolkits like IATK \cite{cordeilIATKImmersiveAnalytics2019} as we found it easier to extend for our purposes. It already supports the Vega-Lite declarative grammar which is very popular in the visualisation community. DXR also uses individual GameObjects for each individual mark, simplifying mesh generation and management.
This came at the cost of rendering performance however, especially when thousands of marks are displayed on the screen. To this end, we made performance improvements to how DXR instantiates and updates its marks and axes by introducing object pooling, especially since marks and axes may be modified multiple times in a morph.
We also added several new visualisation types: choropleth and prism maps, stacked and side-by-side barcharts, and faceted charts (Section~\ref{deimos_sec:example-gallery}). However, as the original DXR implementation does not have support for data transformations like in Vega-Lite, neither does Deimos. This also means that animated transitions involving a time dimension (e.g.\ time varying scatterplots, barchart races) are not supported in Deimos.

\subsection{Code Structure and Pipeline}
Figure~\ref{deimos_fig:deimos-overview} provides an overview of Deimos' structure and how it interacts with our updated version of DXR.
Morph specifications are contained in JSON files that are read by Deimos at initialisation. They can also be refreshed during runtime if the specifications are edited. Deimos interacts with DXR in two main ways. Deimos receives events from DXR whenever a visualisation has been updated, which includes the visualisation specification as an argument. Deimos also sends start and stop function calls to DXR which executes the animated transitions.

Figure~\ref{deimos_fig:deimos-pipeline} provides a high-level overview of the Deimos pipeline in relation to DXR. While it is presented as a linear set of stages, the pipeline can reset or be exited in certain conditions.
First, all morph specifications are read and loaded into Deimos.
Next, whenever a DXR visualisation updates, Deimos is notified via event with the visualisation's specification.
This specification is used to check against all state specifications in the loaded morphs using the rules in Section~\ref{deimos_sss:state-matching-process}.
For any state that has matched, observable streams are created for each signal that is part of the state's transitions, including trigger signals. Observables are created using the UniRx package \cite{kawaiUniRxReactiveExtensions2022} and are composed together where necessary.
When a transition's trigger signal returns true (or if no trigger was specified in the first place), initial and final keyframes are created using the rules in Section~\ref{deimos_sss:keyframe-creation-process}.
These two keyframes, along with other transition parameters such as tweening and staging variables, are sent to the relevant DXR visualisation to start the transition.
When the transition has finished, Deimos stops the transition on the DXR visualisation. This step also updates the visualisation specification to reflect the new changes made by the transition.
Deimos then disposes of all observables related to the transition.
This process then starts anew again, with Deimos finding matching states to see if this newly updated visualisation is eligible for any morphs once more.

While Deimos is designed such that it exists separately from the visualisation framework used, they are still intrinsically linked to each other. Deimos is dependent on the visualisation framework to implement the actual animation and transition. It is also dependent on the grammar and syntax of the visualisations themselves. Therefore, translating Deimos to other visualisation toolkits requires adaptation to support the new declarative grammar, and the toolkit itself needs to support animation between keyframes via interpolation. While it is technically possible to create a middleware to translate visualisation specifications and thus increase modularity, we did not explore this option in this work.

\subsection{XR Interactions}
We use the Mixed Reality Toolkit (MRTK) \cite{microsoftMixedRealityToolkitUnity2022} to enable XR interactions in Deimos. As a result, Deimos can be deployed on a range of platforms including Windows Mixed Reality, Oculus Quest, and HoloLens. However, due to the aforementioned performance limitations when working with large amounts of data, it is recommended to only use Deimos in tethered VR or remote rendering AR setups. Both controller and articulated hand tracking are supported in Deimos in the form of source-based signals (Section~\ref{deimos_sss:source-based-signals}). While Deimos does not support eye gaze or voice input, these can be included in future work.
\section{Example Gallery} \label{deimos_sec:example-gallery}
We present several examples of morphs created with the Deimos grammar. We categorise and describe the examples in three ways, with the first two aligning with the design goals in Section \ref{deimos_sec:design-goals}.
First, we highlight how morphs can be designed to adapt to different visualisation configurations using generic states (DG1), but also allow for bespoke morphs by using specific states in controlled contexts (DG3).
Second, we demonstrate how morphs can be controlled using both embodied (DG2) and non-embodied (DG3) interaction methods.
And third, we provide two scenarios in which Deimos can facilitate the prototyping of different interaction methods.
All examples and their specifications are included in the Deimos Github repository. As such, we do not provide nor go into detail about each example's specification. The project files also contain additional example morphs not described in this paper.

\subsection{Generic vs specific morph examples} \label{deimos_ssc:example-gallery-generic-specific}
In DG1 and DG3, we described a spectrum in which morphs can vary between generic, adapting itself to a range of visualisation configurations, and specific, allowing it to be used in controlled settings.

On the generic end, we present the \textit{3D Barchart Partitioning and Stacking} morph (shown in Figure \ref{deimos_fig:teaser}). It takes a 3D barchart and either partitions it into a 2D faceted barchart, or stacks it into a 2D stacked barchart whenever it touches a surface in the immersive environment. During the transition, it also aligns the visualisation to be parallel against the surface that it had touched. This is an example of a morph involving three states and two transitions in a branch-like structure. The triggers are set up so that the applied transition is based on the angle of contact between the barchart and surface: orthogonal for the faceted barchart, and parallel for the stacked barchart.
Its states are defined such that they only check that the encodings' types are correct (i.e.\ nominal \textit{x} and/or \textit{z}, quantitative \textit{y}) and that it uses cube marks. Through this, so long as a visualisation is a 3D barchart then it can undergo this morph, greatly expanding the range of scenarios it can be used in. JSON path accessors are also used to substitute in the proper field names during runtime (i.e.\ \textit{facetwrap}, \textit{yoffset}).

\begin{figure}
    \centering
    \includegraphics[width=\linewidth]{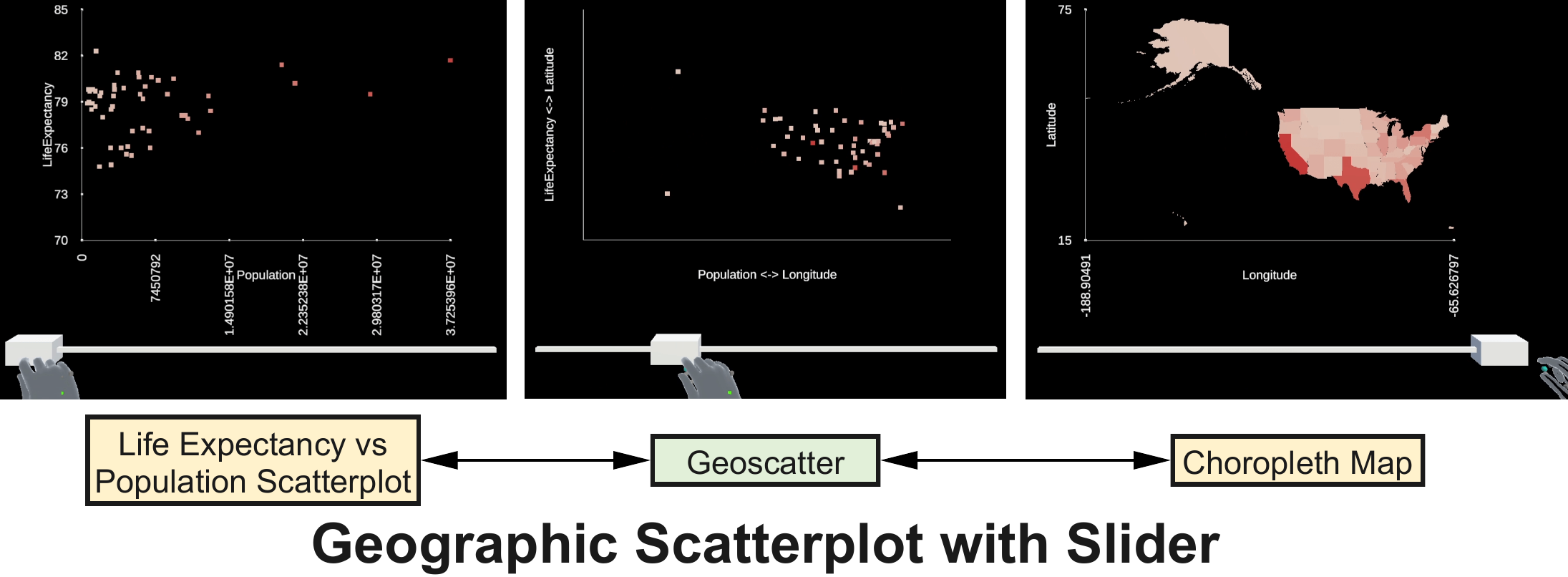}
    \caption{Still images of the \textit{Geographic Scatterplot with Slider} morph, using Unity GameObjects as a slider to control the transition.}
    \Description{A set of still images showing a hand moving a slider from left to right, which causes the points on a 2D scatterplot to move, and then expand into a choropleth map.}
    \label{deimos_fig:deimos-examples-slider}
\end{figure}

\begin{figure*}
    \centering
    \includegraphics[width=0.7\linewidth]{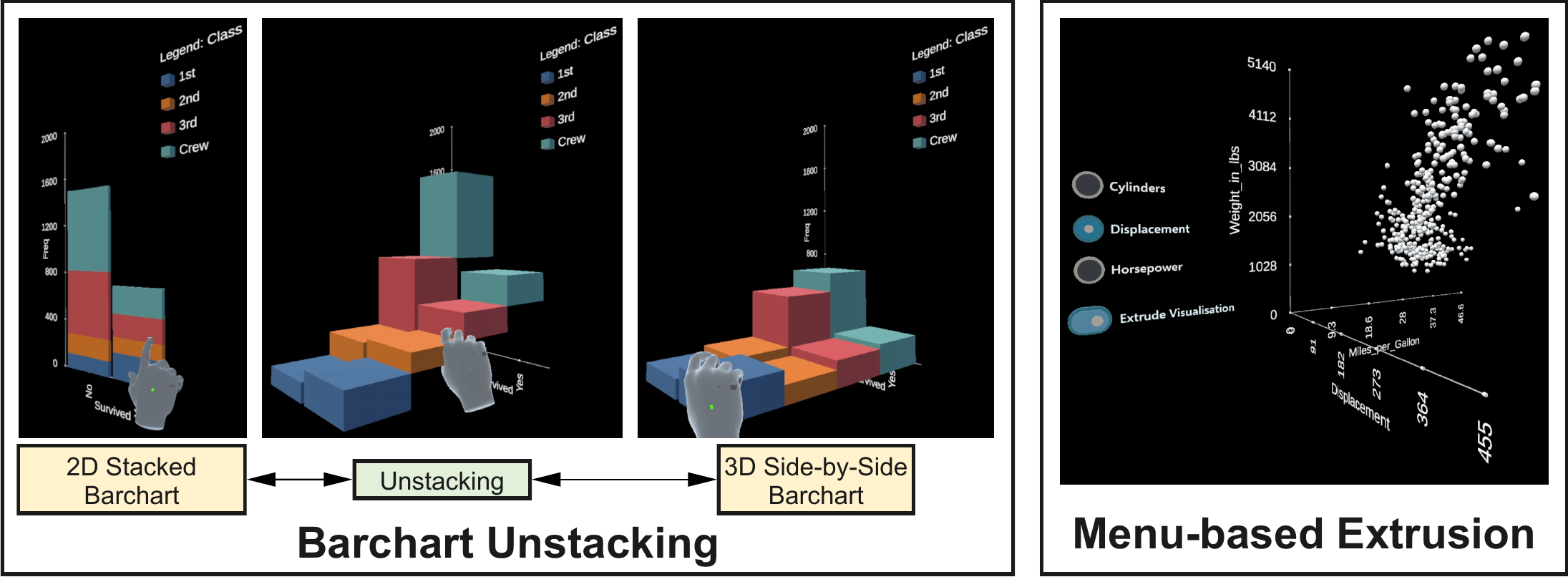}
    \caption{Examples of embodied and non-embodied morphs. Left: Still images of the \textit{Barchart Unstacking} morph, using a ``pinch and pull'' gesture to unstack a 2D barchart into 3D. Right: The result of the \textit{Menu-based Extrusion} morph showing the radial menu and toggle button.}
    \Description{A two-part image. The first part shows a 2D barchart on a surface. A hand grabs onto this barchart and pulls away from it, causing the 2D barchart to extrude out into 3D, and level itself out into a 3D side-by-side barchart. The second part shows a 3D scatterplot. Next to it is a radial menu with the options ``Cylinders'', ``Displacement'', and ``Horsepower'', with Displacement being selected. Below it is a toggle button that is enabled labelled ``Extrude Visualisation''. The 3D scatterplot has a z dimension with the field Displacement.}
    \label{deimos_fig:deimos-examples-embodied}
\end{figure*}

On the other end of the spectrum, the \textit{Geographic Scatterplot with Slider} morph (shown in Figure \ref{deimos_fig:deimos-examples-slider}) demonstrates the use of two predefined states: a scatterplot and a choropleth map. Both of these are explicitly defined using exact encodings and field names (e.g.\ ``Population'', ``LifeExpectancy''). Because of this, only a visualisation with these exact encodings and fields can undergo this morph. A transition connects the two states together, which is controlled using a linear slider represented by a Unity GameObject. A signal accesses the \textit{x} position of this GameObject and uses it as the timing property of the transition.
A morph like this is useful for controlled settings like data-driven storytelling, as the visualisation(s) are all predefined by the author.

\subsection{Embodied vs non-embodied morph examples} \label{deimos_ssc:example-gallery-embodied}
In DG1 and DG3, we described a spectrum in which morphs vary based on the use of embodied vs non-embodied (or WIMP-based) interactions.

On the embodied end, the \textit{Barchart Unstacking} morph uses a ``pinch and pull'' metaphor as the gesture to unstack the bars of a 2D barchart into a side-by-side 3D barchart (shown in Figure \ref{deimos_fig:deimos-examples-embodied} left). To strengthen the metaphor of bars being extruded out into 3D, a condition is added whereby the 2D barchart needs to be positioned against a surface for the morph to be allowed---introducing a contextual requirement to the morph.
To initiate the transition, the user also needs to perform a pinch gesture on the visualisation itself, which is represented by a deictic signal. Other signals calculate the distance between the user's hand and the surface the visualisation is resting against.
The transition uses this distance as its timing property, causing the bars to extrude at the same rate which the user pulls away from them. In this fashion, the user perceives themselves as actually stretching the barchart into 3D, thus resulting in a high level of gestural congruency \cite{johnson-glenbergEmbodiedScienceMixed2017,johnson-glenbergImmersiveVREducation2018}. Of course, this is but one way in which embodied interaction can be achieved, but this approach can be replicated across other morphs to achieve similar styles of extrusion effects.

On the non-embodied end, the \textit{Menu-based Extrusion} morph adds a third spatial dimension to a 2D scatterplot, but does so via an MRTK toggle button \cite{microsoftMixedRealityToolkitUnity2022} (shown in Figure \ref{deimos_fig:deimos-examples-embodied} right). A signal retrieves the state of this toggle button, and will trigger the visualisation when the button is toggled on. This example also demonstrates the use of a radial menu to select the field name of the newly added dimension. A signal retrieves the selected value and substitutes it into the 3D scatterplot state at keyframe creation.
In comparison to the \textit{Barchart Unstacking} morph, this example presents a much simpler and more familiar type of animated transition, albeit in an immersive environment.

\subsection{Prototyping morph interactions} \label{deimos_ssc:example-gallery-prototyping}
Lastly, we demonstrate how the grammar allows for signals to be easily swapped and modified to allow rapid prototyping of different interactions. In terms of the Cognitive Dimensions of Notations \cite{greenCognitiveDimensionsNotations1989}, this corresponds to a low level of \textit{viscosity}.

\begin{figure*}[htb]
    \includegraphics[width=\linewidth]{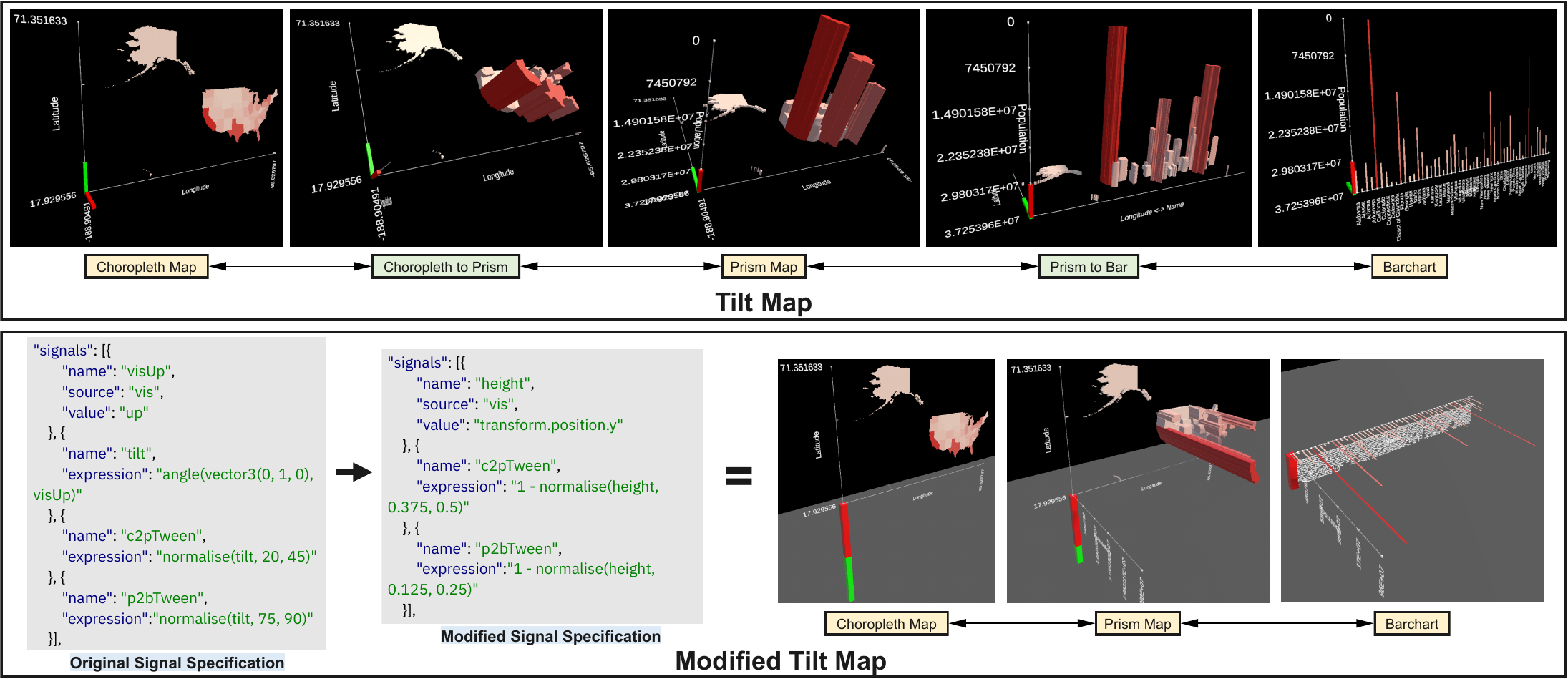}
    \caption{Top: Still images of the \textit{Tilt Map} morph based on Yang et al.\ \cite{yangTiltMapInteractive2020}. A red and green angle bracket is shown to provide rotation cues. Bottom: A modified version of \textit{Tilt Map} showing changes to the signal specification and the resulting morph shown as still images. This example shows tilt being replaced with height. A red and green bar is shown to provide height cues. }
    \Description{A two part image. The first part shows a series of still images of a 2D choropleth map being tilted. As it rotates, it morphs into a prism map. As it rotates even further, it morphs into a 2D barchart. The second part shows a modified version of the first part. Two signal specifications written in JSON are shown which shows the modifications made. Next to it are still images of a choropleth map turning into a prism map, then into a 2D barchart as it is lowered towards the ground.}
    \label{deimos_fig:deimos-examples-tiltmap}
\end{figure*}

In this example, we recreate \textit{Tilt Map} by Yang et al.\ \cite{yangTiltMapInteractive2020} using Deimos (shown in Figure \ref{deimos_fig:deimos-examples-tiltmap} top).  Three states are defined: choropleth map, prism map, and barchart. Two transitions are defined to connect these states linearly. A signal is then created to retrieve the tilt angle of the visualisation relative to the horizontal plane. This tilt angle is then subdivided into two ranges at specific angles using expression signals, that are then used as tweening variables for the two transitions (choropleth to prism, prism to barchart). With this, a visualisation will morph between the different states depending on its tilt.
However, we can easily change the manner which the morph is controlled just by replacing the tilt angle with another source. A straightforward example is to replace it with the height of the visualisation relative to the floor (shown in Figure \ref{deimos_fig:deimos-examples-tiltmap} bottom). The two expression signals which subdivide the range will also need to be updated to the new value ranges. In doing so we turn \textit{Tilt Map} into a so-called ``Height Map'', just by changing a few lines in the morph specification. The result is shown in Figure \ref{deimos_fig:deimos-examples-tiltmap}.

\begin{figure*}[htb]
    \centering
    \includegraphics[width=0.85\linewidth]{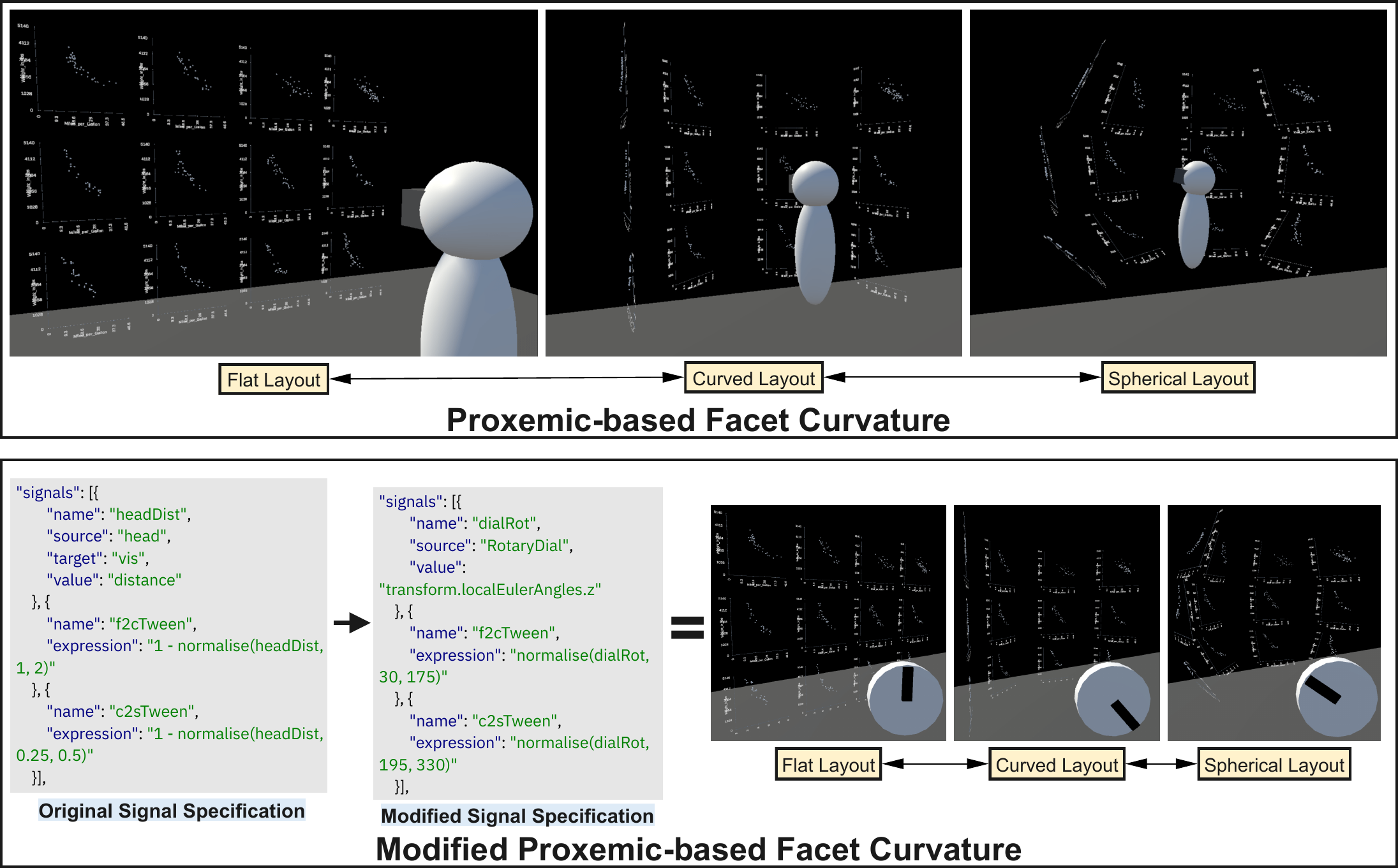}
    \caption{Top: Still images of the \textit{Proxemic-based Facet Curvature} morph, which curves around the user based on the distance between them and the chart. Bottom: A modified version which replaces distance with the rotation of a separate dial object. The changes to the signal specification are shown with the resulting morph shown as still images.}
    \Description{A two part image. The first part shows a series of still images of a faceted chart turning from a flat layout to a curved layout, then into a spherical layout, as a VR avatar approaches it. The second part shows a modified version of the first part. Two signal specifications written in JSON are shown which shows the modifications made. Next to it are still images of the same faceted chart going from flat to curved to spherical, but this time it is controlled using a virtual dial object.}
    \label{deimos_fig:deimos-examples-faceted}
\end{figure*}

Inspired by work on small multiple layouts in immersive environments \cite{liuDesignEvaluationInteractive2020}, we created the \textit{Proxemic-based Facet Curvature} morph (shown in Figure \ref{deimos_fig:deimos-examples-faceted} top). It morphs into a faceted chart between three different layouts: flat, curved, and spherical. These three layouts correspond to three states in the morph, with two transitions connecting them linearly. A signal retrieves the distance between the user's head and the visualisation, with two more signals subdividing the distance into tweening variables (similar to the \textit{Tilt Map} morph). As the user approaches the faceted chart, it begins to wrap around them into a curved layout, and when they are close enough it morphs into an egocentric spherical layout. This effectively makes the chart spatially aware of the user's position.
To demonstrate another method of controlling this morph, we can replace the distance signal with the value of a rotary dial (shown in Figure \ref{deimos_fig:deimos-examples-faceted} bottom). As the user rotates the dial the small multiples curve inwards or outwards. To do so, we create a separate cylinder GameObject in Unity which functions as this dial. We then replace the distance signal with a signal which retrieves the rotation value of the cylinder, and we also update the ranges of the two subdividing signals. This functionally turns the proxemics-based interaction into one involving the manipulation of an external object. This object is currently only virtual, but the concept can be applied to physical objects using either tangible input or motion tracking.

\section{Expert Evaluation} \label{deimos_sec:user-study}
We evaluated Deimos in order to:
(i) determine the ease of use and expressiveness of the grammar;
(ii) get impressions on the concepts introduced in the grammar; and
(iii) generate discussion topics and research directions on the use of animated transitions in immersive environments.

\subsection{Study Design}
We use an approach similar to Zong and Pollock et al.\ \cite{zongAnimatedVegaLiteUnifying2022} by recruiting three developers of immersive analytics grammars and toolkits: Peter Butcher of VRIA \cite{butcherVRIAWebBasedFramework2021}, Philipp Fleck of RagRug \cite{fleckRagRugToolkitSituated2022}, and Ronell Sicat of DXR \cite{sicatDXRToolkitBuilding2019}.
To diversify our participant pool, we also recruited Zeinab Ghaemi of immersive geovisualisation \cite{ghaemiProxemicMapsImmersive2022}, Tica Lin of embedded sports visualisation \cite{linQuestOmniocularsEmbedded2022}, and Jorge Wagner of the VirtualDesk exploration metaphor \cite{wagnerfilhoVirtualDeskComfortableEfficient2018}.
We hoped to learn how Deimos could be positioned within each researcher's respective works.
To minimise learning requirements, we only invited researchers who have experience working with Unity.

The user study was conducted remotely in three sections, repeated for each participant.
First, we conducted a 30-minute introductory session where we explained the goals of the study, demonstrated the examples in Section \ref{deimos_sec:example-gallery}, and went through high-level concepts of the grammar.
Second, we tasked participants to use Deimos unsupervised for at least 2.5 hours. They were given walkthroughs and documentation to learn the grammar, and were encouraged to create their own morphs with some suggestions given to them. This documentation can be found in the Deimos Github repository.
Third, we held a one-hour semi-structured interview based on the aforementioned evaluation goals. 
We asked participants to show us their created morphs, whether they found the overall process easy or difficult, and what parts of the grammar they liked or disliked. For the three participants with toolkit development experience, we also asked how they would retroactively implement animated transitions in their respective toolkits, and if there would be any significant differences compared to Deimos and why. For the other three participants without toolkit development experience, we instead asked how Deimos could be used to support any part of their own research---if at all. However, we allowed the interview to diverge and continue organically, drilling down on any interesting comments participants may have made along the way.
Throughout the study period, we modified the documentation based on participant feedback. While we made bug fixes to Deimos where necessary, we did not add or change any features. Each participant was offered a AU\$150 gift card as compensation for their time.

The interviews were recorded and transcribed. The first author independently performed thematic analysis \cite{braunUsingThematicAnalysis2006} on all six transcriptions, with two other authors doing the same on three transcriptions each. These three authors then discussed and synthesised the main themes together, which form the structure of this section and the following Discussion section.

\subsection{Usability feedback}
We compile participant feedback based on a selection of the most relevant Cognitive Dimensions of Notations \cite{greenCognitiveDimensionsNotations1989}. Rather than using the dimensions as heuristics---a common approach in related works (e.g.\ \cite{satyanarayanDeclarativeInteractionDesign2014, satyanarayanCriticalReflectionsVisualization2019})---we use them from a usability perspective to evaluate the Deimos grammar. However, we provide self-evaluation for certain dimensions where relevant.

\textbf{Error proneness (likelihood of making errors).}
All participants spent the required 2.5 hours using the toolkit, however four of the six spent 7--8 hours using it. The initial reasoning given by most participants was that they enjoyed their time with Deimos and learning how it worked. On further inspection however it was clear that this was in part due to the steep learning curve of the grammar, with Fleck commenting ``I don't feel that three hours are enough.'' We identified several potential causes of this, largely due to grammar's \textit{error proneness}.
First, many participants (Fleck, Ghaemi, Lin, and Wagner) were unfamiliar with the DXR grammar, with even Sicat not having used DXR for three years. As a result, two grammars needed to be learnt, naturally increasing learning time. As the Deimos grammar is intrinsically linked to its visualisation grammar (Section \ref{deimos_ssc:prototype-data-visualisations}), it is apparent that the choice of visualisation package brings not only technical but also notational difficulties.
Second, our documentation assumed full knowledge of Unity and its functions which not all participants had.
Third, the error messages provided by the Deimos prototype were not useful for participants. While the JSON schema validates whether the morph specification is syntactically correct before it is parsed, no check exists for semantic correctness (e.g.\ making sure \textit{name} properties are unique). This has since been corrected in the prototype.
Some participants suggested ways of easing the learning curve. Sicat suggested video tutorials to better explain the grammar, whereas Butcher suggested providing the DXR documentation as pre-reading before the study is even conducted. Interestingly, no participant suggested changes to the grammar itself beyond simple name changes (the terms \textit{signals} and \textit{restrict}). Whether this is due to participants not having had enough time to be exposed to Deimos' advanced features is unclear.

\textbf{Closeness of mapping (closeness to problem domain).}
The lack of grammar changes suggested by participants could be at least partially explained by its \textit{closeness of mapping}. All participants, when asked, had little to no issues understanding how the grammar models the state machine (Figure \ref{deimos_fig:deimos-state-machine-1}). The only participant who raised potential challenges was Fleck, citing the differences between declarative and imperative languages. As Unity primarily uses imperative programming, the shift to a declarative style in Deimos could confuse certain users, particularly when constructing an interaction using signals. We do not believe this to be a major issue however, especially if the immersive visualisations also use a declarative language (e.g.\ DXR \cite{sicatDXRToolkitBuilding2019}, VRIA \cite{butcherVRIAWebBasedFramework2021}).

\textbf{Viscosity (resistance to change).}
After following the walkthroughs, all participants used the same strategy of combining parts of existing examples together to create new morphs to facilitate their learning. For example, Wagner combined the states and transitions of \textit{Tilt Map} example and the signals of the \textit{Proxemic-based Small Multiple Curvature} example to create a rudimentary ``Proxemic Map''. There are only a few examples of participants extending existing examples with completely new components: Sicat remapped the proxemic interaction of the \textit{Proxemic-based Small Multiple Curvature} example with a virtual rotary dial (the same as in Section \ref{deimos_ssc:example-gallery-prototyping}), and Butcher created a stacked barchart to side-by-side barchart morph based on whenever the mouse is clicked. These all demonstrate a low level of \textit{viscosity} within the grammar, as participants were generally able to achieve their goals without issue (minus the aforementioned issues regarding error proneness). The same concept was also described in Section \ref{deimos_ssc:example-gallery-prototyping}.
However, poor error messages introduced viscosity for a few participants. For instance, Lin had tried to create a reduced version of the \textit{3D Barchart Partitioning} example by removing all surface related signals, but the toolkit did not warn her to remove the references to these signals in the states, resulting in errors. This need to keep track of changes in multiple parts of the specification contributes to higher viscosity.

\textbf{Visibility (ability to view components easily).}
Several participants (Fleck, Sicat, and Ghaemi) noted issues relating to the \textit{visibility} of signals in the grammar, primarily due to the large number of possible keywords involved. It was not obvious what options and/or combinations of signals are available without resorting to the documentation, although the JSON schema aided this process. The same participants acknowledged however that this reliance on documentation is fairly normal for toolkits, especially with only a few hours of experience.
From a technical perspective, the Deimos prototype improves visibility by exposing the names of any active morphs and/or transitions on each visualisation, and provides a toggle to print the emitted values of signals to the console for debugging purposes. Further debug messages can also be enabled which show the visualisation specifications of generated keyframes in JSON format. While these features were not explained in the documentation, they were highly useful during the development of Deimos and the creation of our example gallery.
\section{Discussion} \label{deimos_sec:discussion}
This section continues from Section~\ref{deimos_sec:user-study} by summarising the main themes and discussion topics of the semi-structured interviews with our expert participants. We also include several adjacent topics to round out the discussion of immersive morphs---especially in the context of other animated transition grammars.

\textbf{Adaptive morphs.}
While some participants liked the concept of adaptive morphs, others found it getting in the way of their authoring process.
Butcher saw value in adaptive morphs, saying ``I could see why that would be useful, especially if you had a large array of different charts... having it modular just makes sense.''
Wagner thought that ``the premise works well'', but clarified that he would prefer to have ``a [morph] specification for each type of graph'' instead of one hyper-generic morph that applies to all visualisation idioms.
Ghaemi was caught off-guard by this function when her new morph was being applied to other visualisations unintentionally (a result of overly generic states), but was able to reason with modifying the states to ensure that they are more specific.
Fleck and Sicat faced a similar issue, but instead suggested the ability to use an ID to directly target a specific visualisation, skipping the state matching process altogether. This was particularly of relevance to Fleck, where in \textit{RagRug}~\cite{fleckRagRugToolkitSituated2022} ``the user does not create a visualisation [themselves], but the system creates the existing visualisations.''
Overall, participants were able to grasp the concept of adaptive morphs, but it is apparent that their experiences come from the perspective of the morph author. A quantitative evaluation involving data analysis utilising pre-made morphs for practical tasks would be needed to fully evaluate the concept.

\textbf{The purpose of morphs.}
All participants found the examples exciting and interesting, but some had thoughts on their actual purpose. Ghaemi said that morphs are mainly useful when they add or change the data shown, rather than simply remapping encodings (e.g.\ \textit{Stacked Barchart Extrusion} example).
Lin similarly said that she would only use morphs when working with large amounts of data, such as combining proxemics with Shneiderman's mantra~\cite{shneidermanEyesHaveIt1996}, or when working with multiple views, but ``if it's only one smaller data set, and one chart, I probably wouldn't use it to morph between different columns.''
Butcher said that while our example morphs were ``neat and novel'', their animations did not strictly reveal new information, such as a time-varying scatterplot does. 
Therefore, future work should investigate specific use cases for morphs and how morphs may potentially vary between them.

\textbf{Embodied interaction and discoverability.}
The reception to the use of embodied interactions in Deimos (DG2) was positive, but two participants raised discussion topics around their long-term effects. Many of our example morphs use interaction metaphors for embodied interaction (e.g.\ collide with surface, pinch and pull). Sicat expressed concern over the use of these metaphors, saying ``...maybe in my application, pinning to the wall means or does something, and then someone else develops a morph where stick to the wall does something else... that might confuse people... there's no universal rule that says, pinning to the wall should do this.'' When asked if Deimos could play a role in shaping these metaphors, Sicat responded ``I would keep it open for now and just let [researchers] explore'', noting that the field is still not mature yet. He then suggested the use of tooltips to guide users in discovering morphs, especially when conflicting metaphors are used, but stated this is of low priority. In a similar vein, Lin suggested two ways of improving embodied morphs and their discoverability, especially as she had difficulties performing the rotation required for the \textit{3D Barchart Partitioning and Stacking} example. The first was to have the system predict what action the user is about to do, and display the morphs associated with that action in a ``gestural menu'' that the user can select to trigger the morph. The second was to show a preview of the morph while performing the interaction. When asked about the importance of these features, she said that they ``probably [do not] affect the current grammar, because it's more like an assistant towards the completion of certain interactions'', and that they are more like external scripts loaded after the core grammar. Overall, while there are broader implications of the use of embodied interaction in immersive analytics, we see the power in Deimos being used to explore this design space in the long term, rather than immediately prescribing them in this work.

\textbf{GUIs and morph templates.}
Fleck, Sicat, and Ghaemi brought up ideas on how GUIs can be incorporated into Deimos. Fleck suggested the use of data flows in Node-RED to author morph specifications in JSON, similar to how visualisation specifications are created in \textit{RagRug}~\cite{fleckRagRugToolkitSituated2022}. Sicat recalled his own experiences developing DXR's GUI~\cite{sicatDXRToolkitBuilding2019}, noting that a GUI can be useful for non-experts and even end-users to create their own morphs. In a similar vein, Ghaemi said that a GUI would have greatly assisted her learning process with Deimos, citing her lack of experience in both DXR and toolkits in general. However, both participants clarified that the GUI should only cover basic functions, and advanced features should only be accessed in JSON format.
Sicat went on to suggest that the GUI could expose templates for different parts of the grammar that allows users to mix and match and create new morphs, which would be exposed through dropdowns and menus. He compared this idea to how he used the grammar himself, saying ``I went through your examples, copied the morphs and then pasted it into my morphs and then just modified them a bit. So it's kind of [the] same idea, right? Just a different interface. So for non-experts [it] would be super easy.''
Lin suggested something similar except from an interaction perspective, especially as in our included examples ``the interaction you perform is very standardised.'' In other words, a set of template interaction techniques could be provided to accelerate the morph authoring process. This feedback opens many future design possibilities for how a GUI for toolkits like Deimos might look like, especially if it can allow end-users in VR or AR to create and/or modify their own morphs to suit their own needs without needing to write JSON.

\textbf{Inspiration drawn from the toolkit.}
All participants drew interesting comparisons between Deimos and their respective works. Wagner, Ghaemi, and Lin all showed great interest in morphs that transition between 2D and 3D. For Wagner, from the context of his work on VirtualDesk~\cite{wagnerfilhoVirtualDeskComfortableEfficient2018}, said ``it would be very interesting to be able to just snap [3D visualisations] to the desk, and then they project to 2D, which is something that many experts are very comfortable with, but then I could show to them that they can extract [the visualisation] from the desk or from the wall, and try to grab it and look around...'' For Ghaemi whose field is immersive geovisualisation~\cite{ghaemiProxemicMapsImmersive2022}, it was to have the morph directly tied to adding layers to a virtual map, ``[when the] 3D chart collides with the map, the bars could be scattered through the buildings, so I can see the charts on top of the building.'' For Lin, she raised ideas in the context of embedded sports visualisation~\cite{linQuestOmniocularsEmbedded2022}, whereby ``you [can] drag the 2D charts onto a specific player, or maybe drag it onto the court, like the flat ground floor, and then it just suddenly morphs into this heatmap.'' In this sense, rather than a visualisation just morphing between 2D and 3D, it could also morph between being embedded and non-embedded~\cite{willettEmbeddedDataRepresentations2017}.
We then asked whether they could see themselves using Deimos to aid in their research.
Wagner thought that as a proof of concept it would work ``super well'', but cited the poor scalability of the toolkit as a reason against using it.
Ghaemi was receptive, hypothesising that ``the [toolkit] that you have it's, at least, for some of [my ideas], I'm pretty sure that I can implement what I want.'' She also noted that there are no other immersive analytics toolkits that currently enable animated transitions in the manner she desired. Lin said ``there's a high chance that I could use this library to help me prototype some scene to show [sports analysts and coaches].'' After this proof of concept stage however, she would instead develop her own research prototype from the ground up to support specific features such as ``instant data updating''.
Lastly, Butcher said that ``seeing the change in data and understanding what you know, getting something out of it, it's important... certainly not enough attention has been paid to it in the past I don't think, especially in the immersive space.'' He followed this up by saying ``it's definitely something we're going to look at in future for sure, the effect is fantastic.''
While it is expected that not every researcher can make use of the Deimos grammar and the toolkit, our user study clearly demonstrates the significance of this work in generating further research ideas and promoting the study of animated transitions in immersive analytics.

\textbf{Animation authoring paradigms.}
Deimos was originally designed around keyframe animation as its main authoring paradigm. Interestingly, Deimos can technically be seen as having a combination of both keyframe and preset \& templates paradigms. This is arguably a good thing, as Thompson et al.\ \cite{thompsonUnderstandingDesignSpace2020} recommend authoring tools to combine multiple paradigms together to accommodate differences in designers' preferences. In truth, our use of the two paradigms is actually dependent on who is using the morph. In Section~\ref{deimos_ssc:design-goal-1} we described two types of users of Deimos: the person who is creating the morph in a development environment (i.e.\ the ``morph author''), and the person who is actually using the morph in an immersive environment (i.e.\ the ``end-user''). The morph author creates the morph with a keyframe mindset, and the end-user uses the morphs as though they were presets \& templates. Of course, when used for data exploration the VR/AR analyst does not necessarily need to interpret morphs as presets. Much like Data Clips \cite{aminiAuthoringDataDrivenVideos2017} allows for data videos to be created using preset clips however, it is theoretically possible to re-frame Deimos in a similar manner. Morph authors create preset morphs that apply to generic states. End-users then combine these preset morphs together to create linear narratives or non-linear experiences. While this is merely speculative, we believe that future research can consider and further investigate this unique combination of authoring paradigms for animated transitions.

\textbf{Data-driven vs interaction-driven animation.}
Deimos stands apart from other works in the manner in which animations are initiated and viewed by end-users once they are defined. Animations in Animated Vega-Lite \cite{zongAnimatedVegaLiteUnifying2022}, Canis \cite{geCanisHighLevel2020}, Data Animator \cite{thompsonDataAnimatorAuthoring2021} and so on are more data-driven. Specifications are tailored around the intricacies of the loaded dataset, with grammars like Gemini \cite{kimGeminiGrammarRecommender2021} and Gemini{\textsuperscript{2}}~\cite{kimGeminiGeneratingKeyframeOriented2021} even providing recommendation systems to further improve the animations created. Completed animations are then passively viewed by the end-user, with little to no input required to initiate and/or control its playback.
In contrast, Deimos is a more \textit{interaction}-driven grammar. Morph specifications consider not only the change in visual encodings, but also how the user interacts with the system to trigger the morph itself. Completed morphs are then \textit{actively} viewed by the end-user, with them potentially having a high degree of control over the morph's playback and function. This difference is intentional, as immersive environments are inherently more interactive and embodied \cite{marriottImmersiveAnalytics2018} than desktop environments, encouraging users to ``reach out'' and directly manipulate their data. We expect and encourage future research on animations in Immersive Analytics to maintain this interaction-driven mindset---even for presentation and storytelling to better engage and immerse users through interactivity \cite{isenbergImmersiveVisualData2018,leeWatchesAugmentedReality2018}.

\section{Limitations}
Our work naturally has several limitations in regards to the grammar, the technical implementation, and the user study. First, our grammar is built upon several key concepts such as dynamic morphs and embodied interaction. While we aimed to justify these ideas in Section~\ref{deimos_sec:design-goals}, we did not properly evaluate them with actual end-users in VR/AR performing data analysis tasks. Therefore, we cannot confidently say that our approach is quantifiably beneficial for immersive analytics.
Second, our participants were not exposed to all of the functionalities of Deimos. It is certainly possible that there are pain points when using Deimos' advanced functionalities which were not identified due to the limited amount of time participants spent using it. This could include the inability to perform certain embodied gestures with the grammar, or difficulties managing morphs that contain more than 2 or 3 states and/or transitions.
Third, as the grammar is dependent on the visualisation package that it is built upon, many of its limitations are born from DXR \cite{sicatDXRToolkitBuilding2019}. Limitations include the inability to transition between different mark types, lack of runtime data transformations, and overall poor scalability compared to other toolkits like IATK \cite{cordeilIATKImmersiveAnalytics2019} especially when rendering large amounts of data. The inability to transform data (e.g.\ aggregation and filtering) is especially troublesome as it meant that time-varying animations (e.g.\ Gapminder \cite{roslingBestStatsYou2007}) were not considered while designing the grammar, and using certain visualisations in morphs such as barcharts required pre-processing. While we had attempted to add data transformations into DXR ourselves, the challenges in using .NET as a scripting language made it difficult to achieve a syntax remotely equivalent to that of Vega-Lite \cite{satyanarayanVegaLiteGrammarInteractive2017}. We see this as obvious future work, especially as it can allow visualisations to not only morph between encodings, but also between different levels of aggregation, filters, or even different datasets.
\section{Conclusion} \label{deimos_sec:conclusion}
This paper presented Deimos, a grammar and toolkit for prototyping morphs in immersive environments. Morphs are a collection of animated transitions that occur between different defined states, which are triggered and modified by the use of signals. These morphs are dynamically applied to visualisations during runtime, and are capable of leveraging embodied interaction to enable interactive animated transitions. We view Deimos as an initial foray into what a grammar to create embodied animated transitions in immersive environments would look like. While our example gallery and user study demonstrated Deimos' ability to create a wide range of morphs, future work would seek to understand how these morphs are used by actual data analysts and/or audiences of immersive data stories in VR/AR. We also hope that this work fuels greater interest in the use of dynamically morphing embodied visualisations in Immersive Analytics.

\begin{acks}
We thank the six Immersive Analytics researchers who took part in our user study: Peter Butcher, Philipp Fleck, Zeinab Ghaemi, Tica Lin, Ronell Sicat, and Jorge Wagner. We also wish to thank our anonymous reviewers for their valuable feedback, and Jiazhou Liu for his assistance during a pilot study.
\end{acks}

\bibliographystyle{ACM-Reference-Format}
\bibliography{bibliography}

\appendix

\end{document}